\documentclass[journal]{IEEEtran}

\usepackage{cite}
\usepackage{amsmath,amssymb,amsfonts}
\usepackage{algorithmic}
\usepackage{graphicx}
\usepackage{textcomp}
\usepackage{xcolor}
\usepackage{url}
\usepackage[hidelinks]{hyperref}
\usepackage{orcidlink}
\usepackage{soul}
\usepackage{float}
\usepackage{booktabs}
\usepackage[most]{tcolorbox}
\usepackage[font=small]{caption}
\usepackage{enumitem}

\newcommand{\rev}[1]{{\color{black}#1}}
\newcommand{\revb}[1]{{\color{black}#1}}
\newcommand{\finding}[1]{%
  \begin{tcolorbox}[colback=gray!12, colframe=white, coltitle=black, boxrule=0pt, arc=4pt, left=7.15pt, right=7.15pt, top=3pt, bottom=3pt, before skip=4pt, after skip=4pt]
    #1
  \end{tcolorbox}
}
\usepackage{tikz}
\definecolor{darkred}{rgb}{0.6, 0.0, 0.0}
\definecolor{azul}{rgb}{0.0, 0.1882, 0.2863}
\newcommand{\circulovermelho}[1]{%
  \tikz[baseline=(char.base)]{%
    \node[shape=circle, draw=none, fill=darkred, text=white, inner sep=0pt, minimum size=\ht\strutbox] (char) {\scalebox{0.7}{\textbf{#1}}};%
  }%
}
\newcommand{\circuloazul}[1]{%
  \tikz[baseline=(char.base)]{%
    \node[shape=circle, draw=none, fill=azul, text=white, inner sep=0pt, minimum size=\ht\strutbox] (char) {\scalebox{0.7}{\textbf{#1}}};%
  }%
}

\title{Automated Vulnerability Injection in Smart Contracts Using Large Language Models}

\author{Luca~Migliaccio,
        Roberto~Natella~\orcidlink{0000-0003-1084-4824},
        Naghmeh~Ivaki~\orcidlink{0000-0001-8376-6711},
        Nuno~Laranjeiro~\orcidlink{0000-0003-0011-9901},
        and~Marco~Vieira~\orcidlink{0000-0001-5103-8541}%
\thanks{L. Migliaccio is with the Universit\`{a} degli Studi di Napoli Federico II, Naples, Italy (e-mail: lucamigliaccio2000@gmail.com).}%
\thanks{R. Natella is with the Gran Sasso Science Institute (GSSI), L'Aquila, Italy (e-mail: roberto.natella@gssi.it).}%
\thanks{N. Ivaki and N. Laranjeiro are with the University of Coimbra, CISUC/LASI, DEI, Portugal (e-mail: naghmeh@dei.uc.pt; cnl@dei.uc.pt).}%
\thanks{M. Vieira is with the College of Computing and Informatics, University of North Carolina at Charlotte, Charlotte, NC, USA (e-mail: marco.vieira@charlotte.edu).}}

\IEEEtitleabstractindextext{%
\begin{abstract}
    Assessing vulnerability detection tools for smart contracts requires datasets with known ground truth, yet such datasets are scarce and difficult to build by hand. \revb{We propose an approach that uses Large Language Models (LLMs) to automatically inject vulnerabilities into Solidity smart contracts, and demonstrate it in a case study targeting 49 vulnerability types from OpenSCV.} Injected contracts are validated through a multi-step pipeline checking compilation, execution, business logic, and the presence of the intended vulnerability. Applied to real-world contracts from SmartBugs, LLMs generate nearly 1,000 candidate variants; after deduplication and validation, 32 confirmed vulnerable contracts spanning 25 vulnerability types survive (a 16.58\% survival rate). Surviving contracts concentrate in structurally simpler targets and vulnerability types with localized syntactic patterns. We report practical challenges including LLMs' non-determinism and the difficulty of preserving contract semantics. We then use the validated contracts to assess three static analyzers, revealing complementary and incomplete coverage profiles. Results show that LLM-based vulnerability injection is feasible, while exposing key limitations in scalability and diversity.
\end{abstract}

}

\begin{document}
\pagestyle{plain}

\maketitle
\IEEEdisplaynontitleabstractindextext
\IEEEpeerreviewmaketitle
\thispagestyle{plain}

\section{Introduction}

\IEEEPARstart{S}{mart} contracts are a key component of modern blockchain platforms, \rev{where they carry out digital agreements automatically and without a trusted intermediary} in domains such as Decentralized Finance (DeFi) \cite{lin2022survey}, supply chain management, and digital identity \cite{taherdoost2023smartcontracts}. However, their high economic value makes them attractive targets for malicious actors. \rev{Because} smart contracts are immutable once deployed on the blockchain, any vulnerability in the source code can be exploited for profit, potentially causing significant financial losses \cite{nzuva2019smartcontracts}. Consequently, a substantial body of work has focused on developing automated vulnerability detection techniques, including static and dynamic analysis as well as artificial intelligence and \revb{Large Language Model (LLM)} based approaches \cite{vidal2024slr_sc, alsunaidi2025ml_survey}.

A common challenge across all detection approaches is the difficulty of evaluating their effectiveness. \revb{How well these tools are evaluated matters in practice, since developers rely on them as an assurance mechanism before deploying code that cannot be patched, and the confidence a tool deserves depends on how well its detection capabilities have been measured.} Reliable evaluation requires datasets of smart contracts with known vulnerabilities. In practice, such datasets are scarce. Existing datasets are typically small, manually curated, and limited in the range of vulnerability types they cover \cite{boi2024llm_sc_vuln, sheng2025llm_security_survey}. \revb{Furthermore,} they are difficult to extend, as adding new vulnerability types or increasing the number of contracts requires repeating the same manual effort. This lack of large and extensible datasets limits the ability to compare vulnerability detection techniques and assess their real-world applicability. The problem is particularly severe for AI-based approaches, which require large volumes of labeled data to train and generalize effectively.

One way to address this gap is to artificially inject vulnerabilities into safe contracts, producing labeled datasets in which the ground truth is available by construction. Software fault injection has been used in other domains to assess system dependability \cite{natella2016sfi_survey} and security, for instance, through vulnerability injection in web applications \cite{fonseca2009vuln_injection} and through automated tools for large-scale vulnerability data generation in general-purpose software \cite{nong2024vinj}. \revb{Vulnerability injection thus plays the same role for detection tools that fault injection plays in dependability evaluation, supplying a known reference against which their coverage can be judged.}

In the context of smart contracts, the topic remains largely unexplored. \revb{Ghaleb and Pattabiraman \cite{ghaleb2020issta} inject predefined bug patterns to evaluate static analysis tools,} Iuliano et al. \cite{iuliano2025muse} propose a mutation-based approach to inject vulnerabilities into Solidity contracts, while Chu et al. \cite{chu2024sgdl} use generative adversarial networks to produce vulnerability fragments that are then inserted into contracts via abstract syntax tree manipulation. However, these approaches are typically limited to a small set of vulnerability types and lack a structured validation process to verify that the injected vulnerabilities are correct and realistic.

\revb{This paper proposes an approach for the automated injection of vulnerabilities into Solidity smart contracts using LLMs. The approach uses LLMs both to identify which vulnerabilities can be injected into a given contract and to perform the injection, and validates every injected contract before it is accepted into the resulting dataset. We evaluate the approach through a case study driven by the following research questions:}

\begin{description}
    \item[RQ1:] How accurately can LLMs identify the injectability of different vulnerability types into real smart contracts?
    \item[RQ2:] What proportion of vulnerabilities is correctly injected by LLMs, and which vulnerability types are most suitable for injection?
    \item[RQ3:] Are LLM-injected contracts useful to analyze the detection capabilities of existing static analysis tools (SATs)?
\end{description}

\revb{The approach comprises three phases.} Phase~I selects target safe contracts and configures the experimental setup. Phase~II uses one LLM to assess which vulnerability types can be injected into each contract, and a second LLM to perform the injection under predefined constraints. Phase~III validates the injected contracts through compilation, injection verification, business logic verification, and vulnerability confirmation, producing a verified dataset\revb{, which is the final output of the approach. In the case study, we then use this dataset to analyze the detection capabilities of existing SATs}.

\revb{The case study uses} 14 safe Solidity smart contracts drawn from the SmartBugs dataset, spanning diverse application types and structural complexity profiles. Targeting 49 vulnerability types from the OpenSCV taxonomy~\cite{vidal2024openscv}, LLMs generate nearly 1,000 candidate vulnerable variants, which are then passed through a multi-step validation pipeline covering compilation, execution, business logic preservation, and manual vulnerability confirmation.
All code, data, and scripts are publicly available in a replication package~\cite{replication2026}. The main contributions of this work are the following:
\begin{itemize}
    \item \revb{An approach} for LLM-based vulnerability injection into Solidity smart contracts, including multi-step validation that verifies compilation, constraint compliance, business logic integrity, and vulnerability correctness;
    \item \revb{A case study evaluating the approach}, including a characterization of the factors affecting injection success;
    \item A demonstration that the verified dataset can be used to analyze the detection capabilities of SATs, revealing complementary and incomplete coverage profiles;
    \item Practical lessons learned about model non-determinism, validation bottlenecks, injection diversity, and legacy-tool limitations.
\end{itemize}

\revb{In summary, the assessment stage is conservative and narrows the space of injection targets effectively, business logic preservation is the step that candidates are least likely to pass, and only a small fraction survive the full pipeline. Survivors concentrate in structurally simpler contracts and in vulnerability types with localized syntactic patterns. On the validated contracts, the static analysis tools show complementary but incomplete coverage, and several contracts cannot be analyzed at all because the tools fail on legacy Solidity code.}

The remainder of this paper \rev{is organized as follows. Section~II reviews related work on vulnerability detection in smart contracts and on fault and vulnerability injection, and positions our work with respect to it. Section~III \revb{presents the proposed approach and its three phases}. Section~IV \revb{reports a case study that applies the approach to real-world contracts, from contract selection to the validated dataset, and closes with a demonstration that uses the dataset to compare static analysis tools}. Section~V discusses the lessons learned and their practical implications, Section~VI covers the threats to validity, and Section~VII concludes the paper.}

\section{Related Work}
\label{sec:related}

\rev{\subsection{Vulnerability Detection in Smart Contracts}
Smart contracts are exposed to a wide range of vulnerabilities,} from improper access control and unsafe external calls to arithmetic errors and reentrancy. Vidal et al.\ \cite{vidal2024slr_sc} survey more than 100 detection techniques and \rev{show that static analysis is the dominant strategy, with tools such as Mythril, Slither, and Oyente applying symbolic execution, taint analysis, or pattern matching. Dynamic approaches complement static analysis by exercising concrete execution paths, but} no single tool achieves full coverage, and combining \rev{multiple analyzers remains the most effective strategy \cite{vidal2021prdc}. Durieux et al.\ \cite{durieux2020icse} demonstrated this at scale by running nine tools on Ethereum contracts, producing the SmartBugs dataset that has since become a standard resource \cite{ferreira2023smartbugs2}. The vulnerabilities that these tools target are catalogued in the SWC registry, a widely adopted enumeration, and in OpenSCV \cite{vidal2024openscv}, which} integrates SWC, CWE, and DASP into a hierarchical taxonomy.

LLMs have \rev{recently been applied to smart contract security as well. Sun et al.\ \cite{sun2024gptscan} combine GPT with program analysis to detect logic vulnerabilities, Boi et al.\ \cite{boi2024llm_sc_vuln} evaluate how well LLMs detect vulnerabilities directly from Solidity source code, and Sheng et al.\ \cite{sheng2025llm_security_survey} survey LLM-based vulnerability detection across several software domains. These studies show that LLMs identify certain vulnerability patterns with competitive accuracy, but their effectiveness depends on the quality and size of} labeled datasets, which are scarce for smart contracts \cite{alsunaidi2025ml_survey}.

\subsection{Vulnerability and Fault Injection}
Software fault injection has \rev{a long history as a technique for assessing system dependability. Natella et al.\ \cite{natella2016sfi_survey} survey the field and identify the main design choices, among them fault models, injection targets, and workload selection. In the security domain, Fonseca et al.\ \cite{fonseca2009vuln_injection} injected realistic vulnerabilities into web applications to evaluate vulnerability scanners. More recently, Nong et al.\ \cite{nong2024vinj} proposed VinJ, which uses code transformations} to generate large-scale vulnerability data for general-purpose \rev{software.} In the blockchain domain, Vidal et al.\ \cite{vidal2025elusive} injected faults \rev{into smart contracts and showed that many of them elude popular verification tools. Beyond smart contracts, Foster et al.\ \cite{foster2025ach} deployed an LLM-based fault generation system at Meta that produces realistic mutants for Android code, showing that LLMs can generate targeted code mutations at scale.

Building reliable ground truth for smart contracts is itself difficult. Di Angelo et al.\ \cite{diangelo2023groundtruth} show that existing ground truth data suffers from inconsistencies and limited coverage. SolidiFI \cite{ghaleb2020issta} approaches the problem through injection, using predefined bug patterns to seed seven vulnerability types into Solidity contracts, but relies on} fixed operators that do not adapt to contract \rev{structure. Iuliano et al.\ \cite{iuliano2025muse} propose \revb{MuSe, a mutation-based approach that injects six vulnerability types} and does not verify that injected faults preserve business logic. Chu et al.\ \cite{chu2024sgdl} use generative adversarial networks to produce vulnerability fragments inserted through AST manipulation, \revb{and cover seven types. Mutation testing tools for Solidity, such as SuMo~\cite{barboni2022sumo}, also generate faulty contract variants through predefined operators, but the seeded faults are generic programming mistakes for assessing test suites rather than specific vulnerability types.}

\revb{Chen et al.}\ \cite{shen2026forge} proposed FORGE, which puts LLMs to a complementary use, extracting and classifying vulnerabilities from real-world audit reports to build labeled datasets. FORGE produces datasets grounded in real incidents, but the resulting labels depend on the quality of the source reports. Our approach generates vulnerabilities by construction, so we decide the type and the placement of each injected flaw.

\subsection{Relation to Our Earlier Work}
This study builds on our own earlier work. The OpenSCV taxonomy that supplies the vulnerability types we inject was proposed in \cite{vidal2024openscv}, after a systematic literature review of detection techniques for smart contracts \cite{vidal2024slr_sc} and an empirical evaluation of verification tools \cite{vidal2021prdc}. We later analyzed how elusive faults affect blockchain reliability and found that many injected faults escape the tools meant to catch them \cite{vidal2025elusive}, which is what motivates the present work. On the injection side, our design choices follow the fault models and target selection discussed in a survey of software fault injection \cite{natella2016sfi_survey}, and the practice of seeding realistic vulnerabilities to evaluate detection tools, which we applied earlier to web applications \cite{fonseca2009vuln_injection}. \revb{This paper adds an LLM as the injection mechanism and} a validation pipeline that decides whether each generated contract is usable as ground truth.

Compared to this work, we inject with an LLM directly into contract code, adapting to the structure of each contract rather than relying on fixed operators or AST templates. We also reach a wider part of the taxonomy, targeting 49 OpenSCV vulnerability types, and we check every injection} through a four-step pipeline covering compilation, constraint compliance, \rev{business logic} integrity, and vulnerability correctness. \revb{Table~\ref{tab:related_comparison} summarizes the approaches most closely related to ours.}

\begin{table}[tbp]
    \centering
    \caption{\revb{Approaches for building vulnerable smart contract datasets}}
    \label{tab:related_comparison}
    \renewcommand{\arraystretch}{1.2}
    \setlength{\tabcolsep}{3pt}
    \scriptsize
    \revb{
    \begin{tabular}{@{}p{1.1cm}p{2.6cm}p{1.2cm}p{2.9cm}@{}}
        \toprule
        \textbf{Work} & \textbf{Injection mechanism} & \textbf{Vuln.\ types} & \textbf{Validation of generated data} \\
        \midrule
        SolidiFI~\cite{ghaleb2020issta} & Predefined bug patterns inserted as self-contained snippets & 7 & Vulnerable by construction (predefined snippets) \\
        SuMo~\cite{barboni2022sumo} & Generic and Solidity-specific mutation operators (44) & None (generic faults) & Mutants assessed through test execution; vulnerabilities not targeted \\
        SGDL~\cite{chu2024sgdl} & GAN-generated fragments inserted via AST manipulation & 7 & Activation checked through deployment and manual review; business logic not checked \\
        MuSe~\cite{iuliano2025muse} & Pattern-based mutation operators & 6 & Injections checked with a static analyzer; business logic not checked \\
        FORGE~\cite{shen2026forge} & No injection; LLM-based extraction from audit reports & n/a (mined) & Manual assessment of extraction precision and label consistency; depends on the source reports \\
        This work & LLM-based in-place injection, adapted to each contract & 49 targeted (25 injectable) & Four-step pipeline covering compilation, constraint compliance, business logic, and vulnerability correctness \\
        \bottomrule
    \end{tabular}
    }
\end{table}

\section{\revb{Vulnerability Injection Approach}}
\label{sec:approach}

\rev{Fig.~\ref{fig:study_overview}} presents an overview of \revb{the proposed approach}, which is organized in \revb{three} phases:

\begin{figure*}[htbp]
    \centering
    \includegraphics[width=0.7\textwidth]{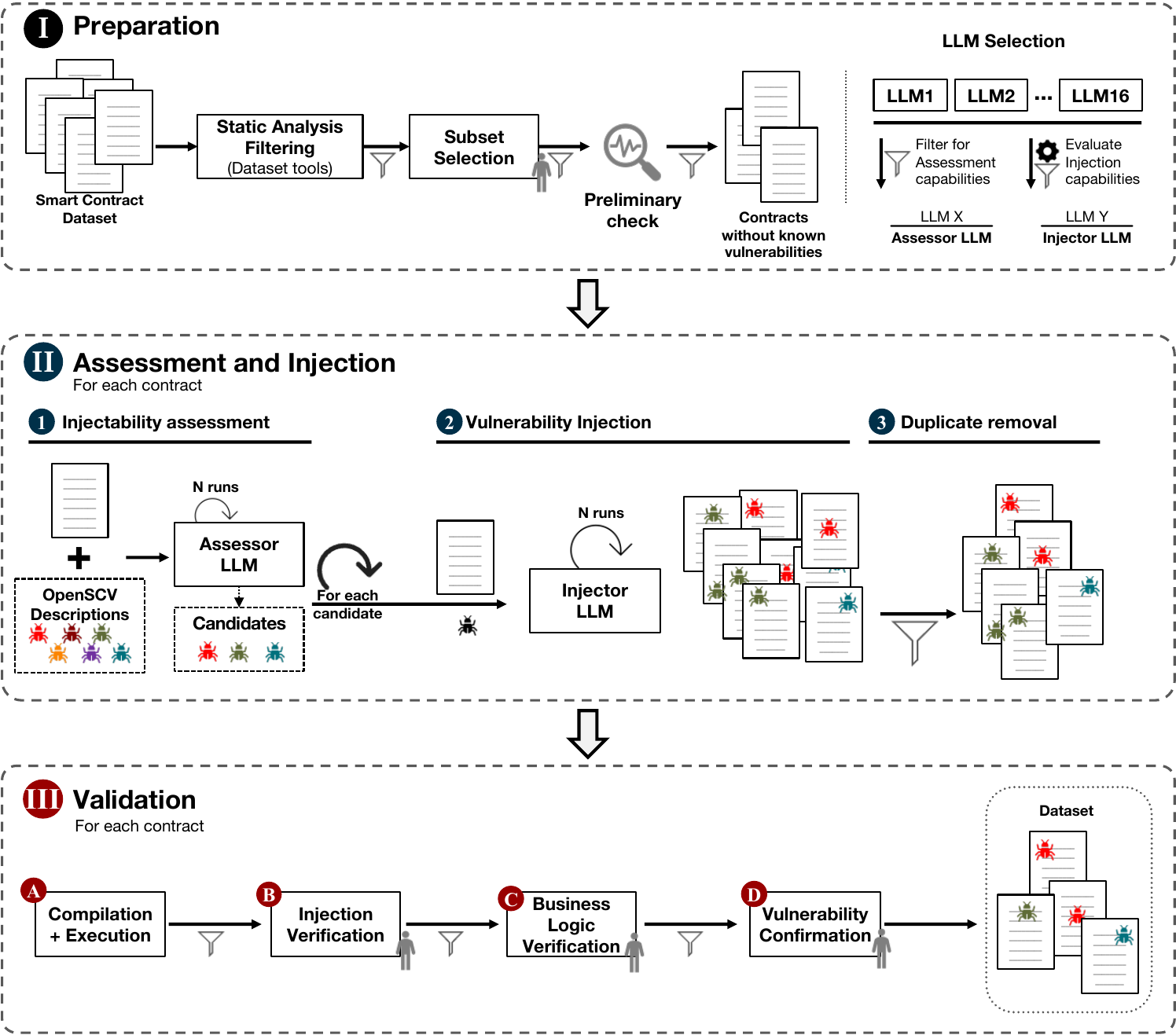}
    \caption{\revb{Overview of the proposed approach}}
    \label{fig:study_overview}
\end{figure*}

\begin{itemize}
    \item \textbf{Phase~I -- Preparation.} Selects target contracts from a large set of real-world contracts, filtering out those with known vulnerabilities and retaining a diverse subset for injection. \revb{It also sets up the experiments, including the choice of the LLMs and the configuration of the injection campaign.}

    \item \textbf{Phase~II -- Assessment and Injection.} Uses \revb{two LLMs, one} for assessing which vulnerability types can plausibly be injected into each contract (Step~1), and another for performing the injection under predefined constraints (Step~2), followed by a duplicate removal step (Step~3).

    \item \textbf{Phase~III -- Validation.} Progressively filters the injected contracts through compilation and execution (Step~A), injection verification (Step~B), business logic verification (Step~C), and vulnerability confirmation (Step~D), producing a verified dataset of vulnerable contracts.
\end{itemize}

\revb{The validated dataset produced in Phase~III is the final output of the approach. Section~\ref{sec:results} presents a case study that applies the approach to real-world contracts and then uses the resulting dataset to analyze the detection capabilities of existing SATs.}


\subsection{Phase I -- Preparation}
We start from a set of smart contracts with no known vulnerabilities. \revb{Candidate contracts are screened with static analysis tools, and only those for which no tool reports a vulnerability are retained.} From this filtered set, a subset of representative contracts is selected to cover diverse application types and code structures.
A final check via SATs confirms the absence of critical vulnerabilities, while minor issues are left in place as they do not affect injection. 
Selected contracts and their complexity profiles are presented in Section~\ref{sec:results}.

\revb{Preparation also covers the setup of the experiments. This includes choosing the LLM used for injectability assessment and the LLM used for injection, defining the vulnerability types to target, and fixing the parameters of the injection campaign, in particular the number of repeated runs per contract--vulnerability pair, which accounts for LLM non-determinism, and the threshold applied to the injectability scores. The concrete choices made in our experiments, covering the selected models and their configuration, the target vulnerability types, and the hardware used, are described in Section~\ref{sec:results}.}

\subsection{Phase~II -- Injectability Assessment \& Injection}
Phase II identifies which vulnerability types can be introduced into each target contract and generates the corresponding vulnerable variants. It consists of three steps: injectability assessment, vulnerability injection, and deduplication.

\textbf{Step \circuloazul{1} Injectability Assessment}. \revb{The assessment LLM, selected in Phase~I, determines, for each target contract, whether each vulnerability type in the target set could realistically be introduced through appropriate modifications. The vulnerability types are taken from the OpenSCV taxonomy \cite{vidal2024openscv}, which provides structured, multi-level vulnerability descriptions well suited for prompt construction, together with paired safe and vulnerable code examples that the injection step reuses. The specific types targeted in our experiments, and the criteria behind their selection, are presented in Section~\ref{sec:results}.}

\begin{figure}[tbp]
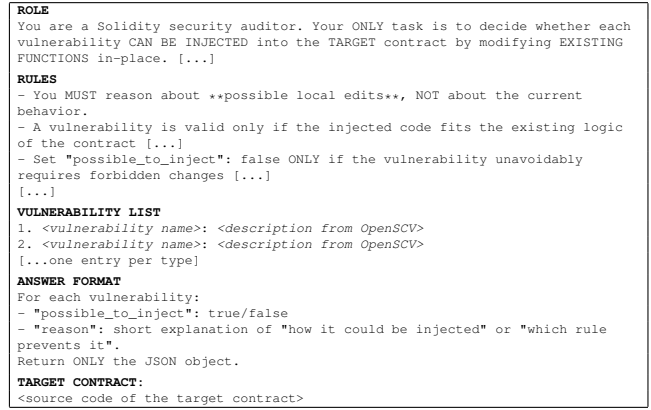

    \centering
    \tiny
    \setlength{\tabcolsep}{3pt}
    \begin{tabular}{|p{0.93\columnwidth}|}
        \hline
        \texttt{\textbf{ROLE}}                                                                                                                                                                           \\
        \texttt{You are a Solidity security auditor. Your ONLY task is to decide whether each vulnerability CAN BE INJECTED into the TARGET contract by modifying EXISTING FUNCTIONS in-place. [\ldots]} \\[2pt]
        \texttt{\textbf{RULES}}                                                                                                                                                                          \\
        \texttt{- You MUST reason about **possible local edits**, NOT about the current behavior.}                                                                                                       \\
        \texttt{- A vulnerability is valid only if the injected code fits the existing logic of the contract [\ldots]}                                                                                   \\
        \texttt{- Set "possible\_to\_inject": false ONLY if the vulnerability unavoidably requires forbidden changes [\ldots]}                                                                           \\
        \texttt{[\ldots]}                                                                                                                                                                                \\[2pt]
        \texttt{\textbf{VULNERABILITY LIST}}                                                                                                                                                             \\
        \texttt{1. \textit{<vulnerability name>}: \textit{<description from OpenSCV>}}                                                                                                                   \\
        \texttt{2. \textit{<vulnerability name>}: \textit{<description from OpenSCV>}}                                                                                                                   \\
        \texttt{\revb{[\ldots one entry per type]}}                                                                                                                                                       \\[2pt]
        \texttt{\textbf{ANSWER FORMAT}}                                                                                                                                                                  \\
        \texttt{For each vulnerability:}                                                                                                                                                                 \\
        \texttt{- "possible\_to\_inject": true/false}                                                                                                                                                    \\
        \texttt{- "reason": short explanation of "how it could be injected" or "which rule prevents it".}                                                                                                \\
        \texttt{Return ONLY the JSON object.}                                                                                                                                                            \\[2pt]
        \texttt{\textbf{TARGET CONTRACT:}}                                                                                                                                                               \\
        \texttt{<source code of the target contract>}                                                                                                                                                    \\
        \hline
    \end{tabular}
    \caption{Structure of the assessment prompt}
    \label{fig:assessment_prompt}
\end{figure}

The assessment prompt is structured in four sections, as shown in \rev{Fig.~\ref{fig:assessment_prompt}:} (i)~a \textit{role} definition that instructs the model to act as a Solidity security auditor; (ii)~a set of \textit{rules} specifying that the model must reason about possible local edits rather than describing the current behavior of the contract; (iii)~a \textit{vulnerability list} containing \revb{one entry per targeted type, with the identifiers} and the textual descriptions provided by OpenSCV; and (iv)~an \textit{answer format} requiring a structured JSON output with a binary \texttt{possible\_to\_inject} field and a brief \texttt{reason} justification for each vulnerability type. This structured role-plus-rules design was preferred over alternatives such as chain-of-thought or few-shot for two reasons: the task requires a decision for \revb{all targeted vulnerability types} in a single call, making few-shot examples impractical due to context length constraints; and the explicit rules provide hard boundaries on the reasoning scope (local edits only) more reliably than emergent CoT reasoning~\cite{white2023prompt_patterns}.

The rules explicitly forbid structural changes such as adding new functions, state variables, or altering the contract's pragma or inheritance. When uncertain, the model is instructed to prefer \texttt{false}, i.e., to predict that the vulnerability cannot be injected into the contract. This design choice reduces false positives at the cost of higher false negatives, which is acceptable in this context because false positives would propagate unrealistic vulnerability types to the injection step.

\rev{Because} LLM outputs are non-deterministic, each evaluation is repeated multiple times\revb{, with the number of repetitions per contract--vulnerability pair defined during preparation}. The results are aggregated into an injectability score, defined as the percentage of runs in which the model predicts that a given vulnerability is injectable:
\begin{equation}\label{eq:injectability_score}
    \small
    \mathit{score}(v, c) = \frac{\mathit{true\_count}(v, c)}{n} \times 100
\end{equation}
where $n$ is the number of repeated evaluations for each contract--vulnerability pair $(v, c)$. This score provides a more stable estimate of a vulnerability type's realistic compatibility with a contract's structure and logic. Only vulnerability types with an injectability score exceeding a predefined threshold are retained for the subsequent injection step.

\textbf{Step \circuloazul{2} Vulnerability Injection}. \revb{The injection LLM, also selected in Phase~I,} is used to generate vulnerable variants for the contract-vulnerability pairs selected during the assessment step. Unlike the assessment prompt, which is shared across all vulnerabilities, the injection prompt is generated individually for each target vulnerability. Each prompt consists of five sections: (i)~a \textit{task description} identifying the specific vulnerability to inject; (ii)~a \textit{vulnerability description} taken from the OpenSCV taxonomy, which defines the target defect in operational terms; (iii)~a set of \textit{constraints} that the model must satisfy; (iv)~a pair of \textit{safe and vulnerable code examples} also taken from OpenSCV, giving the model a concrete pattern to follow; and (v)~the \textit{target contract source code}. This few-shot structure with paired safe/vulnerable examples was selected because it provides the model with a concrete behavioral pattern for the target defect; in our model selection experiments, this strategy consistently outperformed task-only and task-with-vulnerable-example prompting (see Section~\ref{sec:results}).

The constraints are designed to keep injections minimal and realistic. The model must modify exactly one function in one contract, must not add new state variables, contracts, libraries, or functions, and must not introduce any mitigation or defensive logic. The modified line must be annotated with a \texttt{// VULN HERE} comment, which serves as an automated marker for downstream verification. Additionally, the model must preserve the overall logic and ordering of statements, and the resulting contract must remain compilable.

For each eligible contract-vulnerability pair, the injection process is repeated multiple times to capture potential variability in the generated outputs and explore different vulnerable variants compatible with the same objective. \rev{Fig.~\ref{fig:injection_prompt}} shows the general structure of the injection prompt.

\begin{figure}[tbp]
    \centering
    \tiny
    \setlength{\tabcolsep}{3pt}
    \begin{tabular}{|p{0.93\columnwidth}|}
        \hline
        \texttt{\textbf{TASK}}                                                                                                                          \\
        \texttt{Inject a "\textit{<vulnerability name>}" vulnerability into this Solidity contract and mark the new vulnerable line with // VULN HERE.} \\[2pt]
        \texttt{\textbf{TARGET DEFECT}}                                                                                                                 \\
        \texttt{- \textit{<vulnerability name>}: \textit{<description from OpenSCV>} [\ldots]}                                                          \\[2pt]
        \texttt{\textbf{CONSTRAINTS} (must ALL be true):}                                                                                               \\
        \texttt{1. DO NOT add new contracts, libraries [\ldots]}                                                                                        \\
        \texttt{2. DO NOT add any mitigation or defensive logic.}                                                                                       \\
        \texttt{3. Modify exactly ONE function and ONE contract.}                                                                                       \\
        \texttt{4. Your change must introduce the target defect [\ldots]}                                                                               \\
        \texttt{[\ldots] 9. Mark the modified line with: // VULN HERE}                                                                                  \\[2pt]
        \texttt{\textbf{EXAMPLE} (do NOT copy literally):}                                                                                              \\
        \texttt{// SAFE: \textit{<safe code example from OpenSCV>}}                                                                                     \\
        \texttt{// VULNERABLE: \textit{<vulnerable code example from OpenSCV>} // VULN HERE}                                                            \\[2pt]
        \texttt{\textbf{TARGET CONTRACT:}}                                                                                                              \\
        \texttt{<source code of the target contract>}                                                                                                   \\
        \hline
    \end{tabular}
    \caption{Structure of the injection prompt}
    \label{fig:injection_prompt}
\end{figure}

\textbf{Step \circuloazul{3} Deduplication}. \rev{Because} repeated runs of the LLM often produce textually identical outputs, a deduplication step is applied before the validation phase. After normalizing whitespace, two outputs are considered duplicates if they are textually identical. This step prevents the validation phase from being inflated by repeated instances of the same injection pattern and yields a more compact, representative set of candidate vulnerable contracts. The result of Phase~II is a curated set of non-duplicate injected contracts, which are then passed to the multi-step validation pipeline described in Phase~III.

\subsection{Phase~III -- Validation of Injected Smart Contracts}
\label{subsec:Phase2}
Building on the injected contracts obtained in Phase~II, this phase focuses on validating them, verifying that the injected vulnerabilities are correct, consistent with the imposed constraints, and semantically meaningful. The validation process is structured as a four-step pipeline (see Phase~III in \rev{Fig.~\ref{fig:study_overview}).}
Results are recorded using a JSON-based schema that captures build status, manual validation outcomes, and static analysis results. The validation is implemented as a filtering pipeline, in which only contracts that satisfy the constraints of each step are propagated to the next.
\revb{The manual steps (B, C, and D) are performed by a human evaluator, who follows predefined acceptance and rejection criteria, applied consistently across all vulnerability types.} The criteria for each step are described in the following paragraphs, and the complete set, together with annotated accepted and rejected examples, is provided in the replication package~\cite{replication2026}.

\textbf{Step \circulovermelho{A} Compilation and Execution}. Injected contracts are first evaluated for compilability and executability. A contract is considered compilable if it can be successfully processed by the compiler without producing errors. A contract is considered executable if it can be deployed successfully and at least one relevant public or external function can be invoked without causing immediate runtime errors or structural inconsistencies. This step is used as a structural integrity check rather than as a full functional test or exploit demonstration.

\textbf{Step \circulovermelho{B} Injection Verification}. Contracts that pass Step~A are manually inspected to verify that the intended vulnerability has been injected and that the modification complies with the constraints defined in Phase~II. In particular, we verify that the change affects existing code only, does not introduce forbidden structural modifications, and remains syntactically consistent with the original contract. Concretely, the evaluator compares the diff between the original and the injected contract, confirming that at least one modification is present, that none of the injection constraints are violated, and that the modification matches the expected pattern of the target vulnerability.

\textbf{Step \circulovermelho{C} Business Logic Verification}. For contracts that pass injection verification, we assess if the injected change preserves the original business logic. The goal is to confirm that the vulnerability does not alter the intended contract behavior beyond the injected flaw. We separate the injected flaw from unintended semantic drift as follows. A behavioral change is accepted as part of the vulnerability when it is the direct effect of the targeted defect and the contract otherwise behaves as the original on inputs unrelated to the flaw. It is rejected as semantic drift when it alters control flow, roles, execution conditions, or functional behavior beyond the injected weakness, for example removing an unrelated access-control check, changing fund-transfer logic unrelated to the defect, or replacing a function's semantics entirely.

\textbf{Step \circulovermelho{D} Vulnerability Confirmation}. Contracts surviving the previous step are manually analyzed to determine whether the injected flaw corresponds to the target vulnerability, is semantically meaningful in the context of the contract, and is realistically exploitable. This step establishes the final ground truth used in \revb{the remainder of the paper}. We consider a vulnerability realistically exploitable when a reachable transaction sequence triggers the flaw and produces a deviation from the intended behavior that benefits an attacker. The evaluator checks this by simulating the transaction sequence and confirming that the exploit chain is technically feasible. This criterion is applied to all vulnerability types: the expected adversarial effect is derived from each type's OpenSCV specification, and access-control weakening is only one instance of it. \revb{The contracts that pass this step form the validated dataset delivered by the approach.}

\section{\revb{Case Study: Injecting Vulnerabilities into Real-World Smart Contracts}}
\label{sec:results}
\revb{This section presents a case study that applies the proposed approach to real-world smart contracts and reports the main findings.} It covers the contract selection and setup (Phase~I), injectability assessment and outcomes (Phase~II), \revb{and} validation results (Phase~III)\revb{, and closes with a demonstration that uses the resulting dataset to compare SATs}.

\subsection{Phase~I: Contract Selection and Experimental Setup}

The dataset of safe smart contracts was built from the sb-wild subset of the SmartBugs framework \cite{durieux2020icse}, which contains 47,398 real contracts extracted from the Ethereum blockchain. From the existing SmartBugs analysis results, we retained only contracts for which none of the nine integrated SATs reported any vulnerabilities, yielding a set of 2,862 contracts.

\begin{figure}[tbp]
    \centering
    \includegraphics[width=\columnwidth]{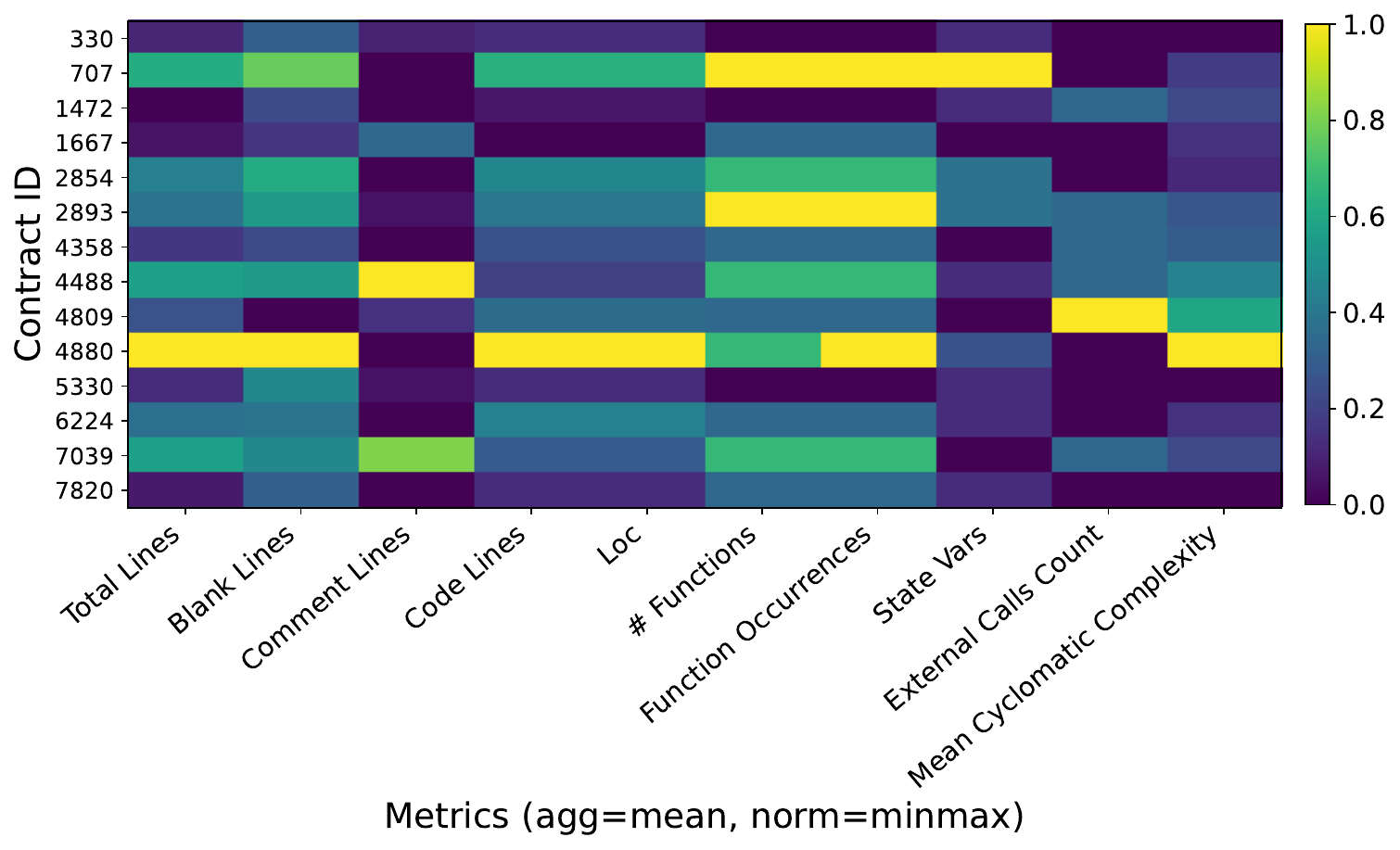}
    \caption{Complexity metrics for the selected contracts}
    \label{fig:contract_metrics}
\end{figure}

From this set, we selected 14 representative contracts covering diverse application \rev{types, including migration,} token, banking, deposit, donation, lottery/game, payment threshold, fund management, merchant, and \rev{termination contracts. We aimed at maximizing diversity across five structural dimensions, namely code size (LOC), control-flow} complexity (CC), number of functions, number of state variables, and external \rev{interactions. These metrics were normalized via min-max scaling to facilitate comparison. A preliminary check was performed with the SATs available in} the Remix IDE (Remix Solidity Analyzer, Slither, and Solhint) \rev{to confirm} the absence of critical vulnerabilities. Of the \rev{14 contracts, 8 were confirmed clean and 2 exhibited minor issues, a weak PRNG and a missing zero-address check, that do not affect the injection process. The remaining 4 could not be fully assessed by static analyzers because of their older Solidity versions and deprecated constructs. We retained them to increase dataset diversity and inspected them manually to gain confidence in the absence of evident vulnerabilities.} This limitation affects only the tool demonstration (Section~\ref{subsec:demonstration}), not \rev{the LLM-based injection.

Fig.~\ref{fig:contract_metrics} shows the normalized complexity profile of each selected contract. The heatmap} confirms that the 14 contracts span a wide \rev{range of structural characteristics,} from minimal contracts with few functions and low \rev{cyclomatic complexity to larger contracts} with extensive external interactions and high control-flow complexity.

All experiments \rev{were conducted on Kaggle, using a P100 GPU with limited VRAM and execution time. This environment constrains model choice but guarantees full reproducibility without proprietary infrastructure. The full experimental pipeline, including code, data, and scripts, is available} in the replication package~\cite{replication2026}. Closed-source models (e.g., GPT-4, Gemini) were excluded from the main pipeline \rev{because they are not freely reproducible and involve per-query costs that limit large-scale experimentation.

We surveyed 16 open-source models spanning a wide range of sizes (66M to 15B parameters), architectures (encoder-only, decoder-only, instruction-tuned, and code-specialized), and training data (general-purpose corpora, multilingual code, and Solidity-specific datasets). From this set, we kept the models compatible with the Kaggle P100 hardware, excluding those that exceed the 16GB VRAM limit of the GPU and those not designed for code, and retaining the ones} relevant to code understanding (assessment) or code generation (injection).

For \rev{the assessment phase, which requires understanding contract structure and semantics to predict whether a vulnerability can be meaningfully injected, we selected Qwen2.5-Coder for its specialization in code analysis. Among the surveyed models compatible with our hardware, it was the strongest code-specialized option available, judged by its results on HumanEval and MBPP, and at 14B parameters it was also the largest code-specialized model that fits within our hardware constraints. Using results on standard code suites} as a proxy for code reasoning capability is consistent with established practice in LLM selection~\cite{sheng2025llm_security_survey}.

For \rev{the injection phase, we evaluated five candidate models, Meta-Llama-3~8B, Qwen3-1.7B-Base, Qwen-2.5-7B-Instruct, Codepy-3B-Base, and NextCoder-7B, on a subset of Solidity~0.8.x contracts using a single vulnerability type (Integer Underflow/Overflow). We chose this setting because Solidity~0.8.x includes built-in overflow protection, so these vulnerabilities are not naturally present and must be explicitly introduced. Each model was tested} under three prompting strategies (task-only, task with vulnerable examples, and task with both vulnerable and safe examples), \rev{with 10 runs per configuration. We selected Meta-Llama-3 because it consistently achieved the highest injection success rates across prompting strategies. This selection} criterion rests on a single vulnerability type and a \rev{subset of contracts, so the relative ranking of models may differ for other defect types. A broader comparative evaluation was not feasible under the hardware limits of the Kaggle P100 environment, and we treat this choice as a pragmatic decision whose impact we discuss in Section~\ref{sec:threats}. Both models were run with quantization to fit within the available resources and used the default generation parameters of llama.cpp, without raising temperature or other sampling settings. We made this choice deliberately, because the injection task leaves little room and we preferred reproducible outputs over greater generation diversity. Table~\ref{tab:llms_used} summarizes the final model} selection.

\begin{table}[tbp]
    \centering
    \caption{LLMs used in the experimental pipeline}
    \label{tab:llms_used}
    \renewcommand{\arraystretch}{1.2}
    \setlength{\tabcolsep}{4pt}
    \resizebox{0.4\textwidth}{!}{
        \begin{tabular}{lclccl}
            \toprule
            \textbf{Model} & \textbf{Size} & \textbf{Type}    & \textbf{Quant.} & \textbf{Context} & \textbf{Role} \\
            \midrule
            Qwen2.5-Coder  & 14B           & Code-specialized & Q4\_0           & 8192             & Assessment    \\
            Meta-Llama-3   & 8B            & General-purpose  & Q8\_0           & 8192             & Injection     \\
            \bottomrule
        \end{tabular}
    }
\end{table}

We consider 49 \rev{vulnerability types from the OpenSCV taxonomy (Table~\ref{tab:openscv_swc}), selected according to two criteria. First, we included only types with a corresponding SWC code \revb{(SWC, Smart Contract Weakness Classification, is a registry of weakness types for Ethereum smart contracts, analogous to the CWE registry for general software)}, because SWC identifiers are widely recognized by both LLMs and SATs, which allows a} direct comparison between injected vulnerabilities and tool \rev{outputs. Second, the number of types was bounded by the LLMs' context length, as including the full OpenSCV taxonomy together with the target contract and the task instructions would exceed the maximum input size, so we retained the subset that fits within the available context window. For} the assessment phase, each contract--vulnerability pair was evaluated 10 times \rev{to account for LLM} non-determinism, and only types with an injectability score (Eq.~\ref{eq:injectability_score}) above 50\% \rev{were retained, so that injectability rests on a majority of the runs. For the injection phase, each eligible contract--vulnerability} pair was processed 10 times under the predefined \rev{injection} constraints. \revb{The demonstration of the dataset (Section~\ref{subsec:demonstration}) uses the same three SATs employed in the preliminary contract check, namely Remix Solidity Analyzer, Slither, and Solhint.}

\begin{table*}[tbp]
    \centering
    \caption{OpenSCV vulnerability types}
    \label{tab:openscv_swc}
    \renewcommand{\arraystretch}{1.0}
    \setlength{\tabcolsep}{2.5pt}
    \scriptsize
    \resizebox{\textwidth}{!}{
        \revb{\begin{tabular}{llcllcllc}
            \toprule
            \textbf{ID} & \textbf{Vulnerability Name}                & \textbf{SWC} & \textbf{ID} & \textbf{Vulnerability Name}                    & \textbf{SWC} & \textbf{ID} & \textbf{Vulnerability Name}                & \textbf{SWC} \\
            \midrule
            1.1.1       & Unsafe Credit Transfer                     & 107          & 5.5         & Wrong Class Inheritance Order                  & 125          & 6.1.6       & Transfer Recip.\ Dep.\ on Trans.\ Order    & 114          \\
            1.1.2       & Unsafe System State Changes                & 107          & 5.6.2       & Function Return Type Mismatch                  & 127          & 6.2.2       & Extraneous Input Validation                & 123          \\
            1.3.1       & Improper Check of Ext.\ Call Return Val.\  & 104          & 5.7.2       & No Effect Code Execution                       & 135          & 7.1.1       & Integer Underflow                          & 101          \\
            1.3.2       & Improper Exc.\ Handling of Ext.\ Calls     & 113          & 5.7.3       & Unused Variables                               & 131          & 7.1.2       & Integer Overflow                           & 101          \\
            1.4         & Improper Locking during Ext.\ Calls        & 132          & 5.8.1       & Undet.\ Program Version Prevalence             & 103          & 7.2.1       & Divide by Zero                             & 101          \\
            1.6         & Delegatecall to Untrusted Callee           & 112          & 5.8.2       & Outdated Compiler Version                      & 102          & 7.2.2       & Integer Division                           & 101          \\
            2.1.2       & Improper Exc.\ Handling in a Loop          & 128          & 5.8.3       & Use of Deprecated Functions                    & 111          & 7.3.1       & Truncation Bugs                            & 101          \\
            2.1.3       & Incorrect Revert Impl.\ in a Loop          & 126          & 5.10.1      & Wrong Function Modifier                        & 100          & 7.3.2       & Signedness Bugs                            & 101          \\
            3.1         & Improper Gas Requirements Checking         & 126          & 5.10.3      & Missing Visibility Mod.\ in Var.\ Decl.\       & 108          & 8.1.1       & Wrong Caller Identification                & 115          \\
            3.2         & Call with Hardcoded Gas Amount             & 134          & 5.12.1      & Use of Same Var./Func.\ Name in Inh.\ Contr.\  & 119          & 8.1.3       & Missing Verif.\ for Program Termination    & 106          \\
            4.2         & Unprotected Transfer Value                 & 105          & 5.13.2      & Write to Arbitrary Storage Location            & 124          & 8.2.1       & Exposed Private Data                       & 136          \\
            5.1         & Bad Randomness                             & 120          & 5.15        & Typographical Error                            & 129          & 8.3.1       & Incorrect Verif.\ of Crypto.\ Signature    & 117          \\
            5.2.1       & Missing Constructor                        & 118          & 6.1.1       & Incorrect Use of Event Blk.\ Var.\ for Time    & 116          & 8.3.2       & Improper Check against Sig.\ Replay Att.\  & 121          \\
            5.2.2       & Wrong Constructor Name                     & 118          & 6.1.2       & Incorrect Function Call Order                  & 114          & 8.3.3       & Improper Authenticity Check                & 122          \\
            5.2.4       & Uninitialized Storage Variables            & 109          & 6.1.3       & Improper Locking                               & 132          & 8.3.4       & Incorrect Argument Encoding                & 133          \\
            5.4.2       & Wrong Selection of Guard Function          & 110          & 6.1.4       & Transfer Pre-Cond.\ Dep.\ on Trans.\ Order     & 114          &             &                                            &              \\
            5.4.3       & Function Call with Wrong Arguments         & 130          & 6.1.5       & Transfer Amount Dep.\ on Trans.\ Order         & 114          &             &                                            &              \\
            \bottomrule
        \end{tabular}}
    }
\end{table*}

\subsection{Phase~II: Assessment and Injection}

This section covers injectability assessment (Step~1), vulnerability injection (Step~2), and deduplication (Step~3).

\subsubsection{Step~1: Injectability Assessment}

\begin{figure}[tbp]
    \centering
    \includegraphics[width=\columnwidth]{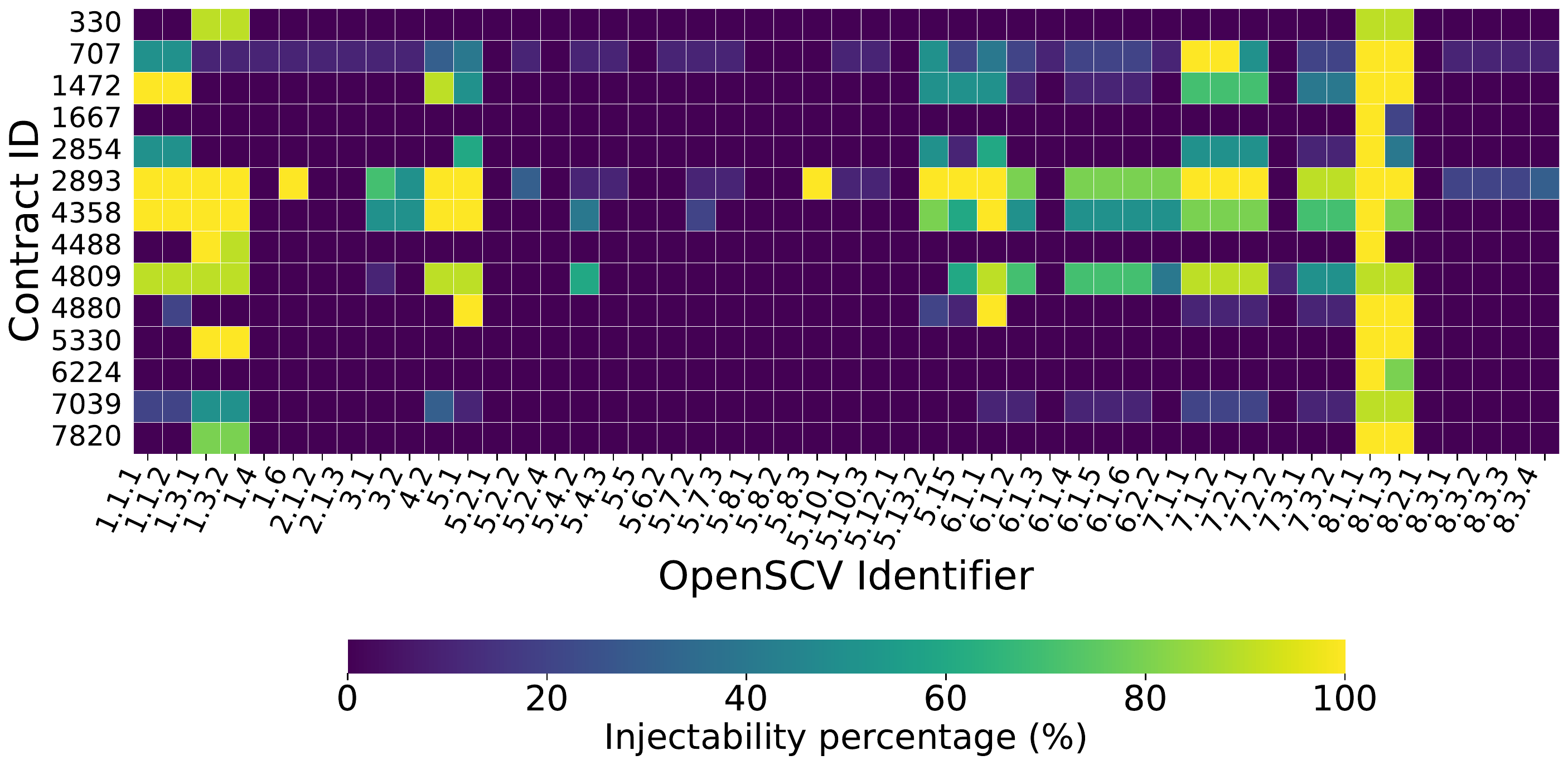}
    \caption{Heatmap of injectability predictions}
    \label{fig:injectability_heatmap}
\end{figure}

Figure \ref{fig:injectability_heatmap} reports the resulting injectability heatmap. The results show that injectability is highly uneven across contract--vulnerability pairs. Only a subset of combinations is consistently judged as injectable, while most pairs receive low or null scores. The global distribution of injectability scores confirms this pattern: scores cluster near 0\% or 100\%, with few intermediate values, indicating that the model produces consistent rather than uncertain predictions. The results also suggest that the model differentiates between vulnerability types that can be realistically introduced through local edits and those that would require structural changes incompatible with the imposed constraints.

\begin{figure}[tbp]
    \centering
    \includegraphics[width=\columnwidth]{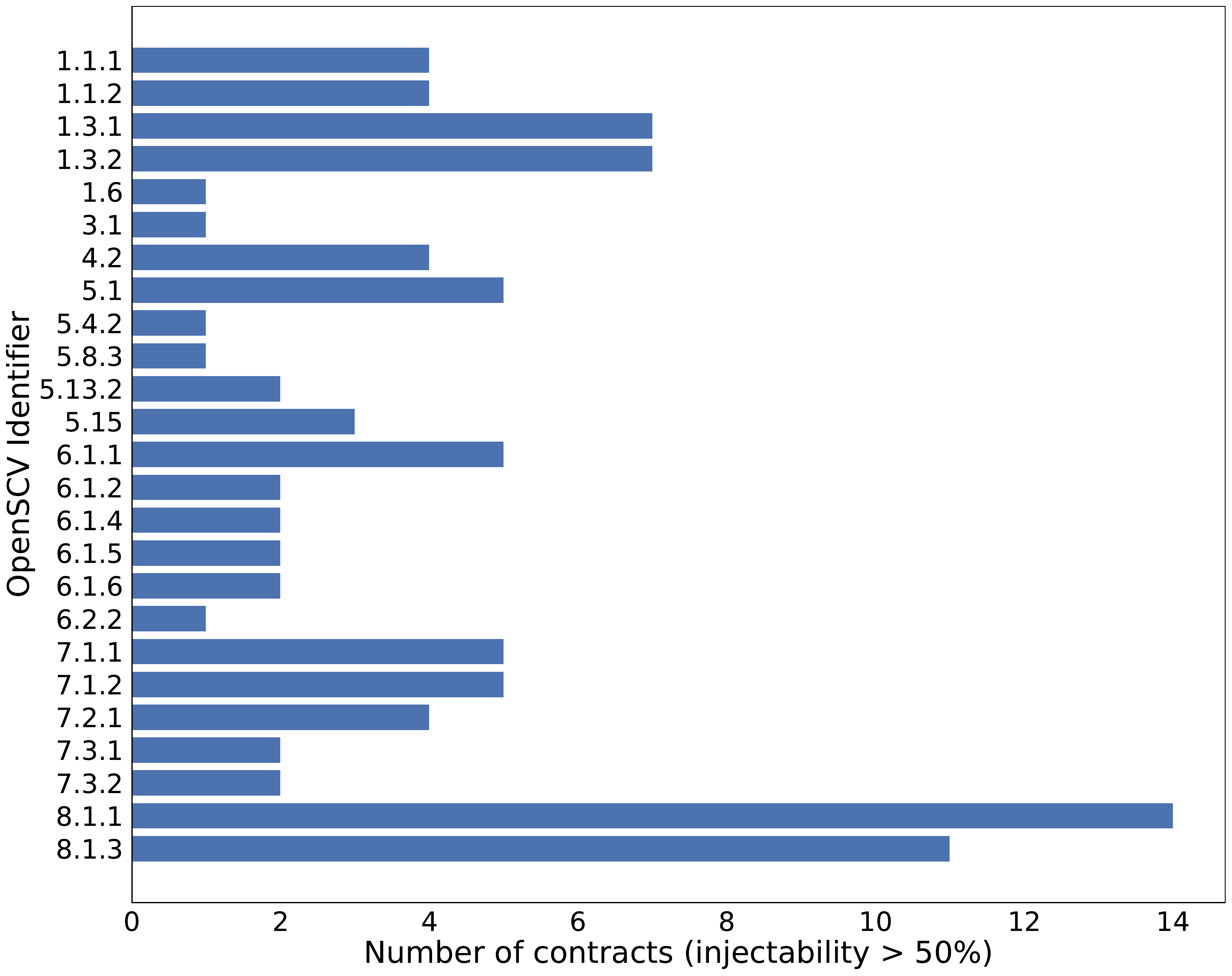}
    \caption{High-injectability per vulnerability type}
    \label{fig:injectability_per_vunerability}
\end{figure}

Of the 49 types initially considered, only 25 satisfy the 50\% injectability threshold for at least one contract and are therefore retained for the subsequent injection step. \rev{Fig.~\ref{fig:injectability_per_vunerability}}
shows the high-injectability vulnerability types. The results suggest that the space of possible injections is substantially smaller than the full vulnerability taxonomy and is strongly constrained by the structure and semantics of the target contracts. Obviously, the number of contracts influences which types reach the threshold, and a larger, more diverse set of contracts might allow additional types to pass. Within the scope of this study, however, the 25 retained types represent the injectable space as observed across the 14 selected contracts.

\begin{figure}[tbp]
    \centering
    \includegraphics[width=\columnwidth]{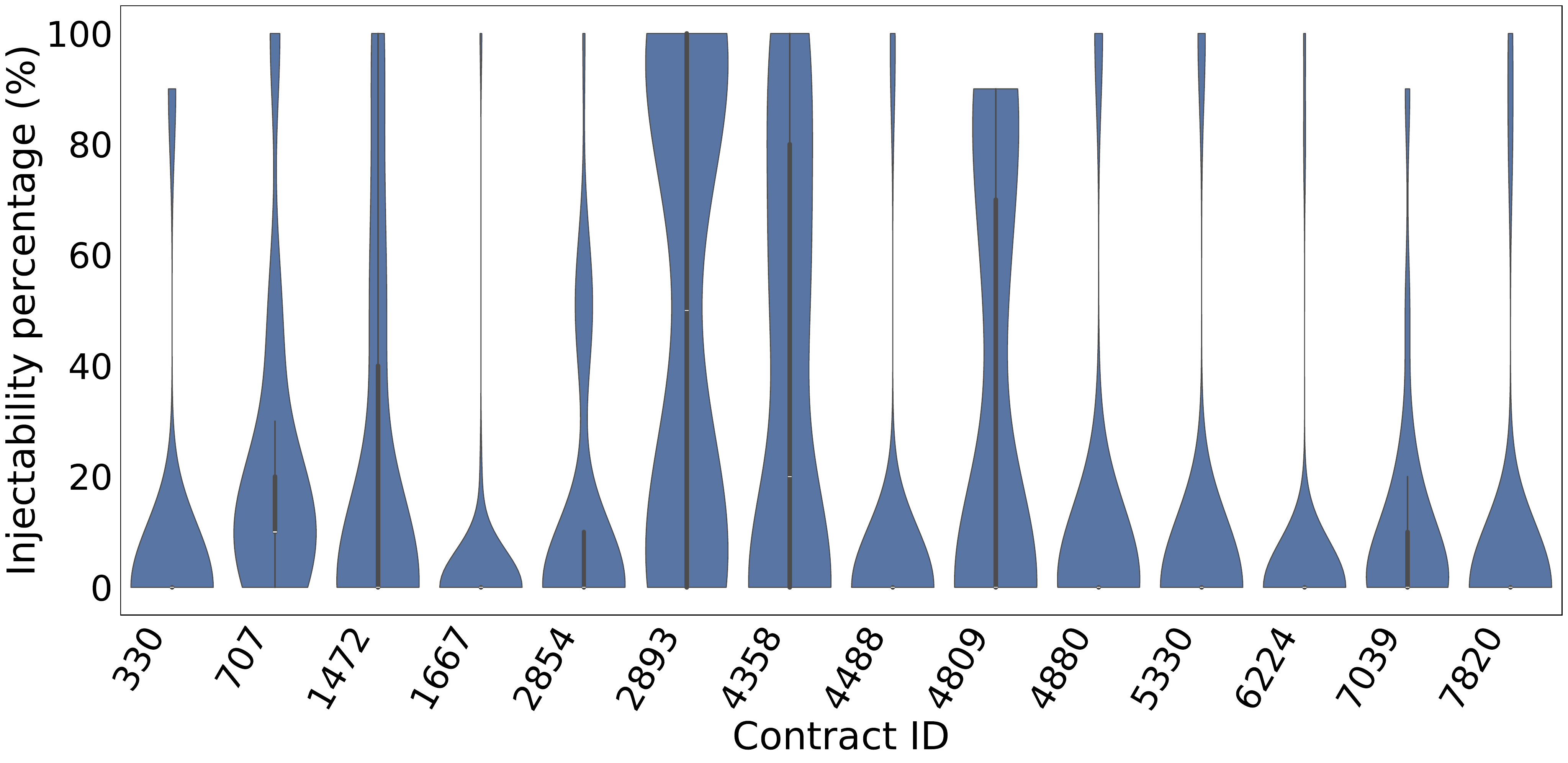}
    \caption{Distribution of injectability scores per contract}
    \label{fig:violin_per_contract}
\end{figure}

The distribution of injectability scores also shows variability across contracts. \rev{Fig.~\ref{fig:violin_per_contract}} shows the per-contract distribution: some contracts exhibit a wide range of injectable vulnerability types, with scores frequently reaching high values, whereas others appear structurally rigid and support only a limited number of plausible injections. Overall, these results suggest that vulnerability injectability is not uniformly distributed, but rather emerges from the interaction between the target contract structure and the specific vulnerability pattern to be introduced.

\begin{figure}[tbp]
    \centering
    \includegraphics[width=0.5\columnwidth]{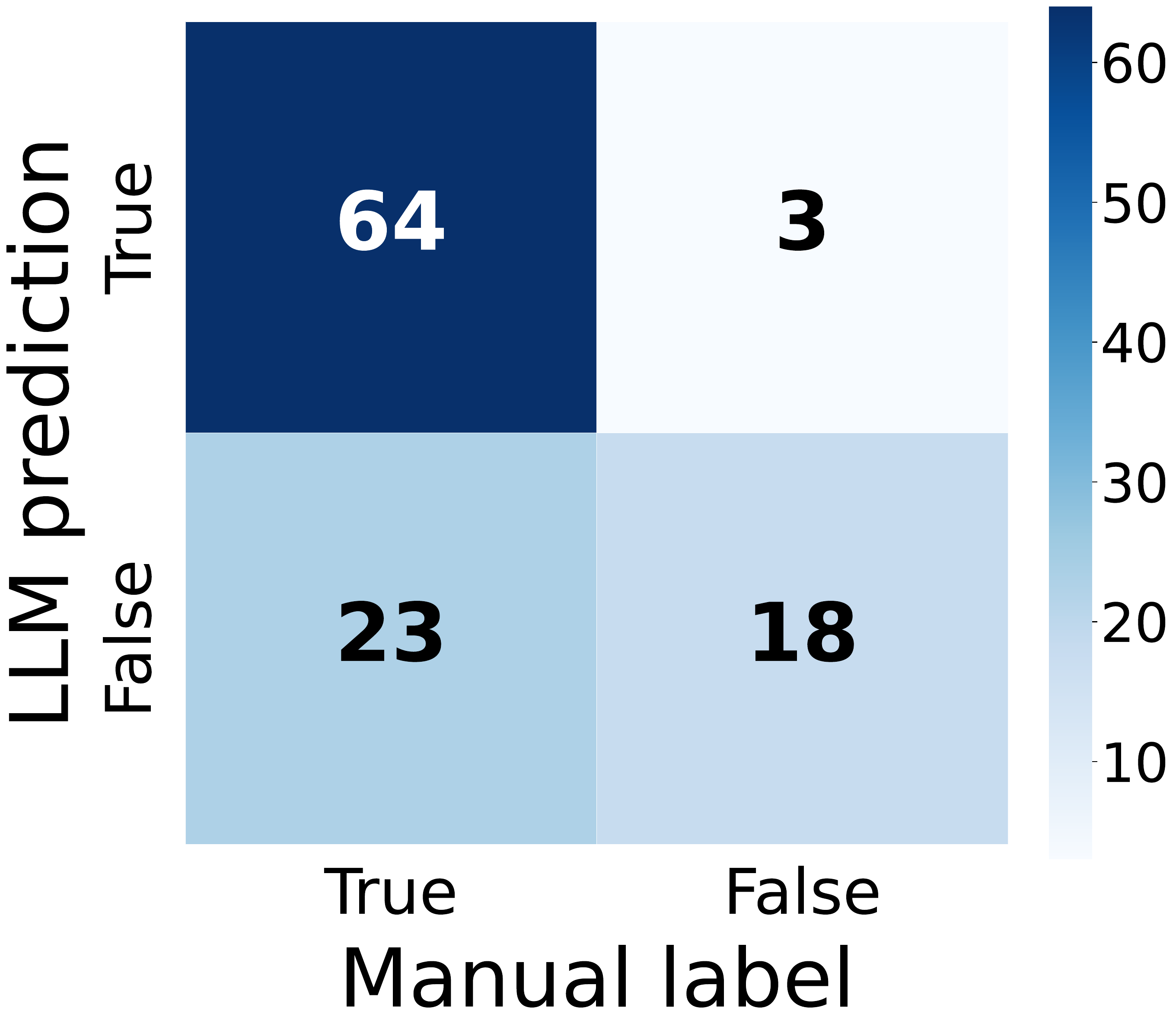}
    \caption{Confusion matrix of assessment decisions}
    \label{fig:confusion_matrix_manual_validation}
\end{figure}

To assess the reliability of the LLM-based assessment, we manually validated a representative sample of 108 decisions across contracts, vulnerabilities, and repeated runs; the sample was stratified to cover all vulnerability types with at least one positive prediction, with decisions drawn from at least two different contracts per type and balanced across positive and negative outcomes. The analysis shows that the model behaves conservatively, with relatively few false positives and more false negatives (Figure \ref{fig:confusion_matrix_manual_validation}). This is a desirable aspect of our pipeline, \rev{because} false positives would propagate unrealistic vulnerability types into the injection phase, whereas false negatives primarily reduce coverage. Manual validation further shows that the model performs better on vulnerabilities with clear, localized syntactic patterns, but struggles with those requiring deeper semantic reasoning or structural preconditions.
These observations indicate that the assessment phase effectively reduces the search space for injection by identifying a narrower set of contract–vulnerability pairs more likely to yield realistic and constraint-compliant results.

\begin{figure}[tbp]
    \centering
    \includegraphics[width=\columnwidth]{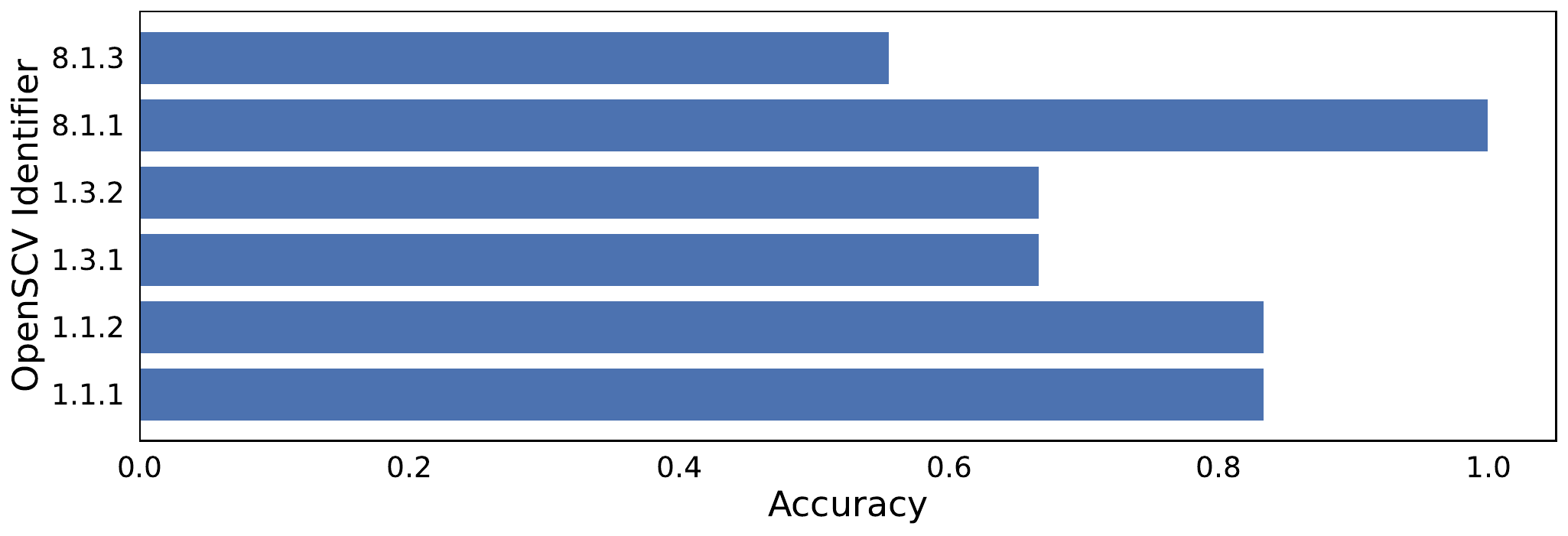}
    \caption{LLM prediction accuracy per vulnerability type}
    \label{fig:accuracy_per_vuln}
\end{figure}

\rev{Fig.~\ref{fig:accuracy_per_vuln}} shows the prediction accuracy broken down by vulnerability type, on a subset of types. The model achieves near-perfect accuracy on types with clear syntactic signatures, such as Wrong Caller Identification (8.1.1), while accuracy drops for types that require deeper semantic reasoning, such as Missing Verification for Program Termination (8.1.3). This confirms that the assessment model is most reliable when the injection pattern can be identified through surface-level code features, and less so when the decision depends on understanding the complete contract logic.

\finding{\textbf{Answer to RQ1:} Manual validation of 108 assessment decisions shows that the LLM achieves 75.9\% accuracy, with high precision (95.5\%) but lower recall (73.6\%). The model produces very few false positives (3 out of 108), meaning it rarely predicts injectability where none exists, but it misses about a quarter of injectable types (23 false negatives). These figures are based on a sample of 108 out of 6,860 total decisions (1.57\%), so they should be interpreted as indicative rather than definitive estimates; the 95\% Wilson confidence interval for accuracy on this sample is approximately [67.3\%, 83.2\%]. Accuracy varies across vulnerability types: types with clear syntactic signatures (e.g., Wrong Caller Identification) reach near-perfect accuracy, while types requiring semantic reasoning show lower reliability. Of 49 types, 25 pass the 50\% injectability threshold for at least one contract.}

\subsubsection{Step~2: Vulnerability Injection}

Building on the vulnerability types selected during the assessment step, the injection phase uses Meta-Llama-3 to generate vulnerable variants of the target contracts. Injection is performed only for vulnerability--contract pairs with an injectability score exceeding 50\%, thereby restricting the generation process to cases predicted to be injectable by the assessment step. For each selected pair, the model is queried multiple times under the predefined injection constraints. As a result, the raw generation process produces many vulnerable candidate contracts, but repeated executions often yield textually identical outputs, making duplicate removal necessary before proceeding to the validation pipeline. Overall, the injection phase generated 997 contracts.

\subsubsection{Step~3: Duplicate Removal}

Following injection, duplicate removal is applied to generated contracts before validation. After normalizing whitespace, contracts are considered duplicates if textually identical. The goal is to retain only distinct vulnerable variants per contract–vulnerability pair, avoiding inflated validation from repeated injection patterns.

After removing duplicates, the number of unique injected contracts decreased to 193, representing an approximately 80\% reduction. For each vulnerability--contract pair $(v,c)$, the number of duplicate generations is defined as $D_{v,c} = N_{v,c} - K_{v,c}$, where $N_{v,c}$ is the number of model executions and $K_{v,c}$ is the number of distinct contract versions generated. Thus, $D_{v,c} / N_{v,c}$ quantifies the degree of injection stabilization. The high redundancy shows that the LLM frequently converges toward recurring transformation patterns rather than exploring a wide variety of distinct solutions, and that the raw number of generations substantially overestimates the effective diversity of the injected output.

\rev{Because} duplicates are identified via text comparison after whitespace normalization, semantically equivalent injections with different formatting (e.g., variable names or comments) are treated as distinct, while textually identical outputs that may affect different execution paths are merged. The count of 193 non-duplicate contracts is therefore an approximation of the true semantic diversity of the generated injections.

\begin{figure}[tbp]
    \centering
    \includegraphics[width=\columnwidth]{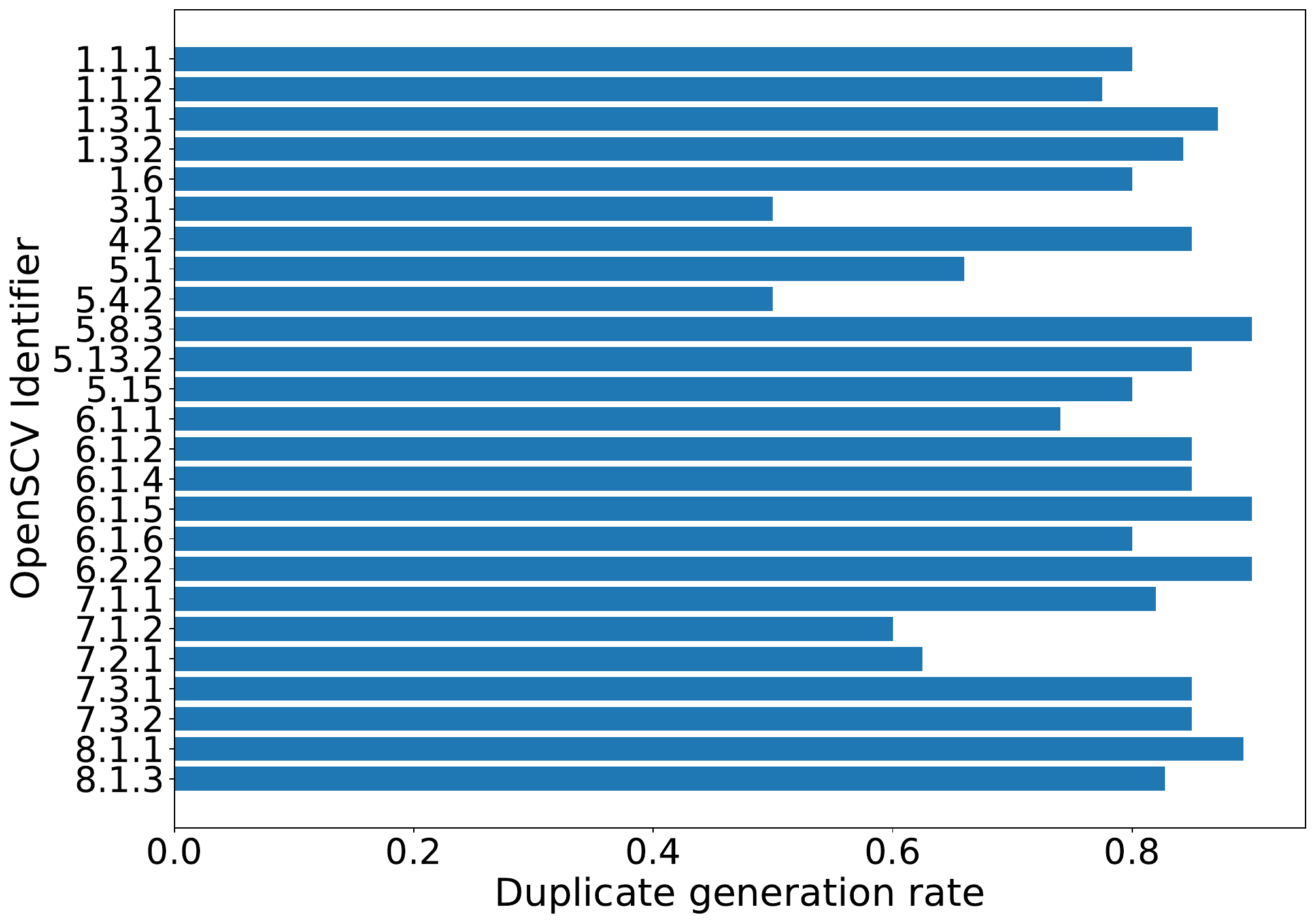}
    \caption{Duplicated LLM-generated contracts per vulnerability}
    \label{fig:dup_per_vuln}
\end{figure}

Figure \ref{fig:dup_per_vuln} shows the ratio of duplicates across vulnerability types. Duplication rates are similar across types, with all categories producing many repeated outputs. This suggests that the model relies on a small set of canonical injection patterns, even for injectable types. As a result, the number of distinct ways a vulnerability can be introduced remains limited.

A different effect can be observed at the contract level, where some targets produce more duplicated generations than others. Contracts with more rigid logic or fewer suitable injection points tend to induce stronger convergence in the model outputs. These findings support two observations: LLM-based injection is strongly shaped by the interaction between vulnerability type and contract structure, and duplicate removal is not merely a data-cleaning step but a necessary phase for estimating the actual diversity of generated contracts.
To further analyze diversity after duplicate removal, we examine the distribution of non-duplicate contracts across vulnerability--contract pairs. This allows evaluating whether the surviving injections are concentrated in a small subset of pairs or distributed across the search space.

\begin{figure}[tbp]
    \centering
    \includegraphics[width=\columnwidth]{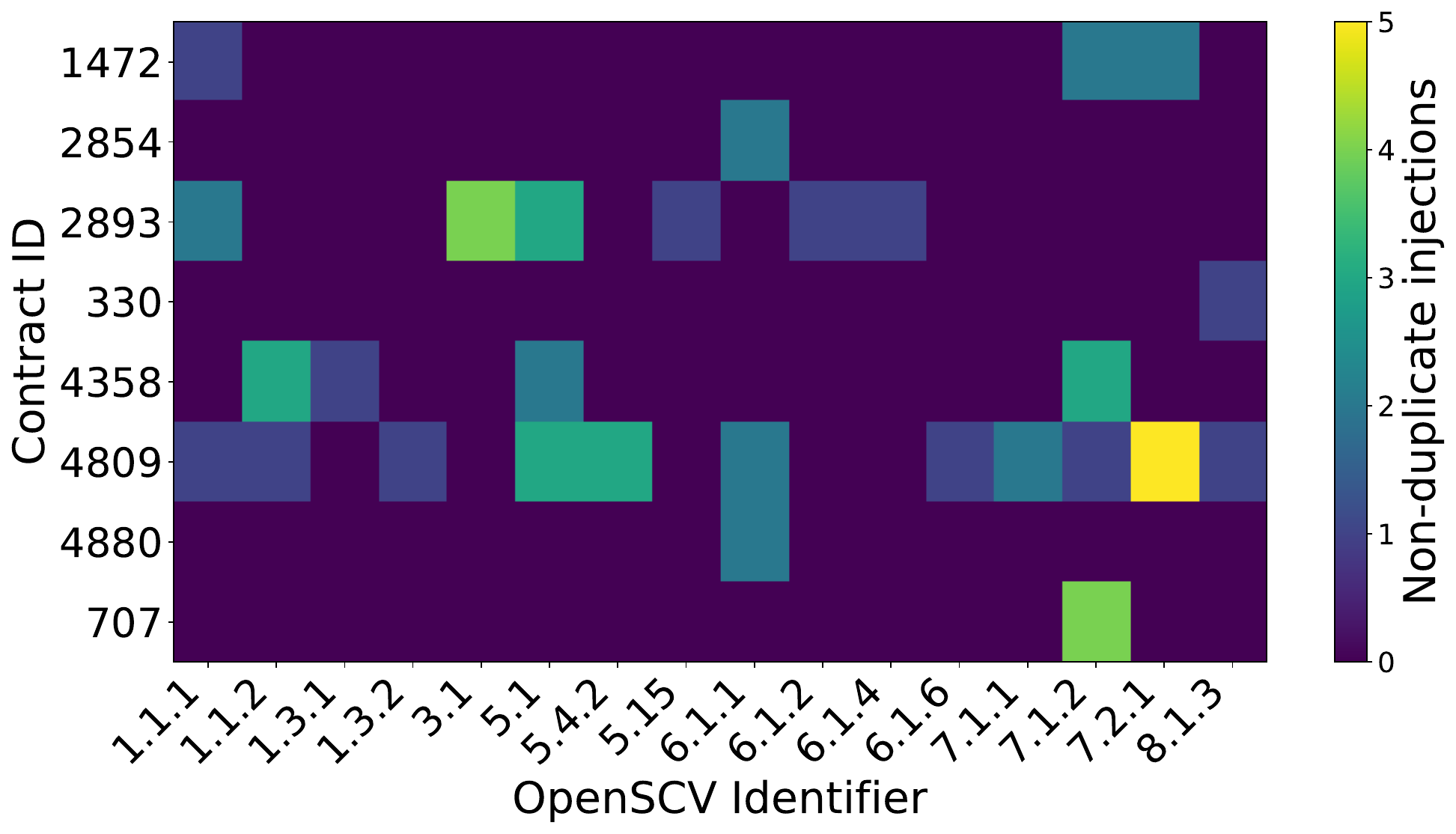}
    \caption{Non-duplicate contracts per contract--vulnerability pair}
    \label{fig:non_dup_per_pair}
\end{figure}

Figure \ref{fig:non_dup_per_pair} reports the number of distinct injected contracts for each contract--vulnerability pair after duplicate removal. Although 14 contracts were initially selected, this heatmap shows only 8 because the remaining contracts resulted in zero non-duplicate injections across all vulnerability types and were excluded to improve readability. The heatmap confirms that diversity is highly uneven: only a subset of pairs yields multiple distinct injected variants, while many others produce either a single unique output or no valid distinct variant at all. This suggests that diversity is not an inherent property of the generation process, but rather depends on the presence of suitable structural hooks in the contract and on the flexibility of the corresponding vulnerability pattern.

The injection results show that the generation phase is effective at producing candidate vulnerable contracts, but also reveal a strong tendency for the model to converge on repeated solutions. The effective output of Phase~II should therefore not be measured in terms of raw generations, but in terms of distinct and semantically meaningful injected variants. The resulting set of 193 contracts constitutes the actual input to the validation pipeline analyzed in Phase~III.

\subsection{Phase~III: Validation}

\subsubsection{Validation Pipeline (Steps A--D)}

Phase~III evaluates the contracts generated in Phase~II through a multi-step validation pipeline. \revb{The manual steps were performed by a single evaluator with experience in Solidity smart contract security; the implications of this setting are discussed in Section~\ref{sec:threats}.} Starting from the 193 non-duplicate contracts, the pipeline progressively filters them based on compilability and executability, injection correctness, preservation of business logic, and final vulnerability validity.
Table~\ref{tab:pipeline_summary} shows the progressive reduction across validation steps. Of the 193 contracts, 150 pass Step~A (Compilation and Execution), 89 pass Step~B (Injection Verification), 44 pass Step~C (Business Logic Verification), and only 32 remain after Step~D (Vulnerability Confirmation), for a final survival rate of 16.58\%.

\begin{table}[tbp]
    \centering
    \caption{Injected contracts surviving each validation step}
    \label{tab:pipeline_summary}
    \renewcommand{\arraystretch}{1.1}
    \setlength{\tabcolsep}{5pt}
    \scriptsize
    \resizebox{0.22\textwidth}{!}{
        \begin{tabular}{lcc}
            \toprule
            \textbf{Step}   & \textbf{Remaining} & \textbf{\% of Initial} \\
            \midrule
            Initial Dataset & 193                & 100.00                 \\
            Step A          & 150                & 77.72                  \\
            Step B          & 89                 & 46.11                  \\
            Step C          & 44                 & 22.80                  \\
            Step D          & 32                 & 16.58                  \\
            \bottomrule
        \end{tabular}
    }
\end{table}

\begin{figure}[tbp]
    \centering
    \includegraphics[width=\columnwidth]{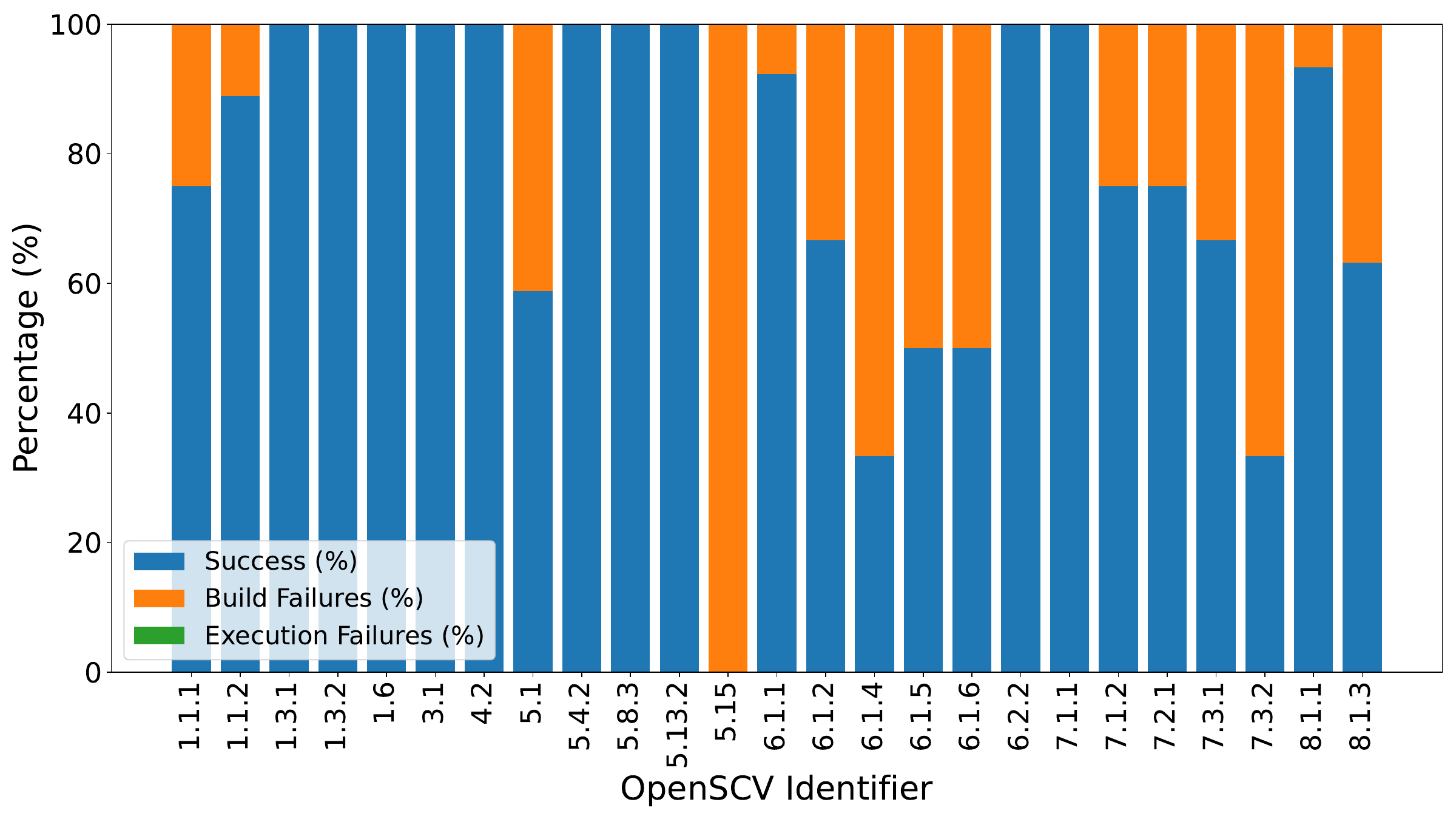}
    \caption{Compilation and execution per vulnerability category}
    \label{fig:step_a_breakdown}
\end{figure}

The results indicate that LLMs are capable of generating verified vulnerable contracts, but multiple filtering steps are required to distinguish formally modified contracts from semantically valid and exploitable vulnerabilities. The largest \revb{relative} reduction occurs from Step~B to Step~C, where survivors drop from 89 to 44, suggesting that preserving business logic while introducing a vulnerability is one of the most challenging aspects of the injection process.
Step~A shows that most injected contracts remain structurally valid, with 77.72\% being compilable and executable. \rev{Fig.~\ref{fig:step_a_breakdown}} breaks down the outcome of Step~A by vulnerability category, showing the proportion of contracts that compile and execute successfully, fail at compilation, or fail at execution. Execution failures are negligible compared to compilation failures across all types. Legacy Solidity versions are a substantial source of these Step~A failures. Of the 43 contracts that fail this step, 26 fail because the injector introduces constructs unsupported by the contract's compiler (most often \texttt{abi.encodePacked} in pre-0.4.24 code), and 40 occur on contracts using Solidity 0.3.x--0.4.x. This is the same root cause behind the static analysis failures reported in Section~\ref{subsec:demonstration} and discussed in Section~\ref{sec:lessons}.

Step~B filters out contracts in which the intended vulnerability is not correctly injected or constraints are violated. The transition from 150 to 89 highlights that constraint-compliant injection remains challenging even when injectability is predicted in Phase~II. Step~C confirms that many syntactically correct injections introduce unintended semantic changes, with three recurring failure patterns: removal or weakening of access control checks (e.g., allowing transfers without sufficient funds); alteration of fund transfer logic (e.g., storage updates unrelated to the injected vulnerability), and complete replacement of function semantics (e.g., a \emph{Kill} function that no longer terminates a contract). Step~D identifies contracts for which the injected vulnerability is present, meaningful, and realistically exploitable (e.g., the injection indeed weakens access control as required by the specification of the vulnerability type). Access-control weakening is used here only as an illustrative case; the exploitability criterion defined in Section~\ref{subsec:Phase2} (Step~D) is applied uniformly across all vulnerability types, with the expected adversarial effect taken from each type's OpenSCV specification. These 32 contracts constitute the ground truth used in the comparison with SATs.

\subsubsection{Factors Affecting Injection Survival}

\begin{figure}[tbp]
    \centering
    \includegraphics[width=\columnwidth]{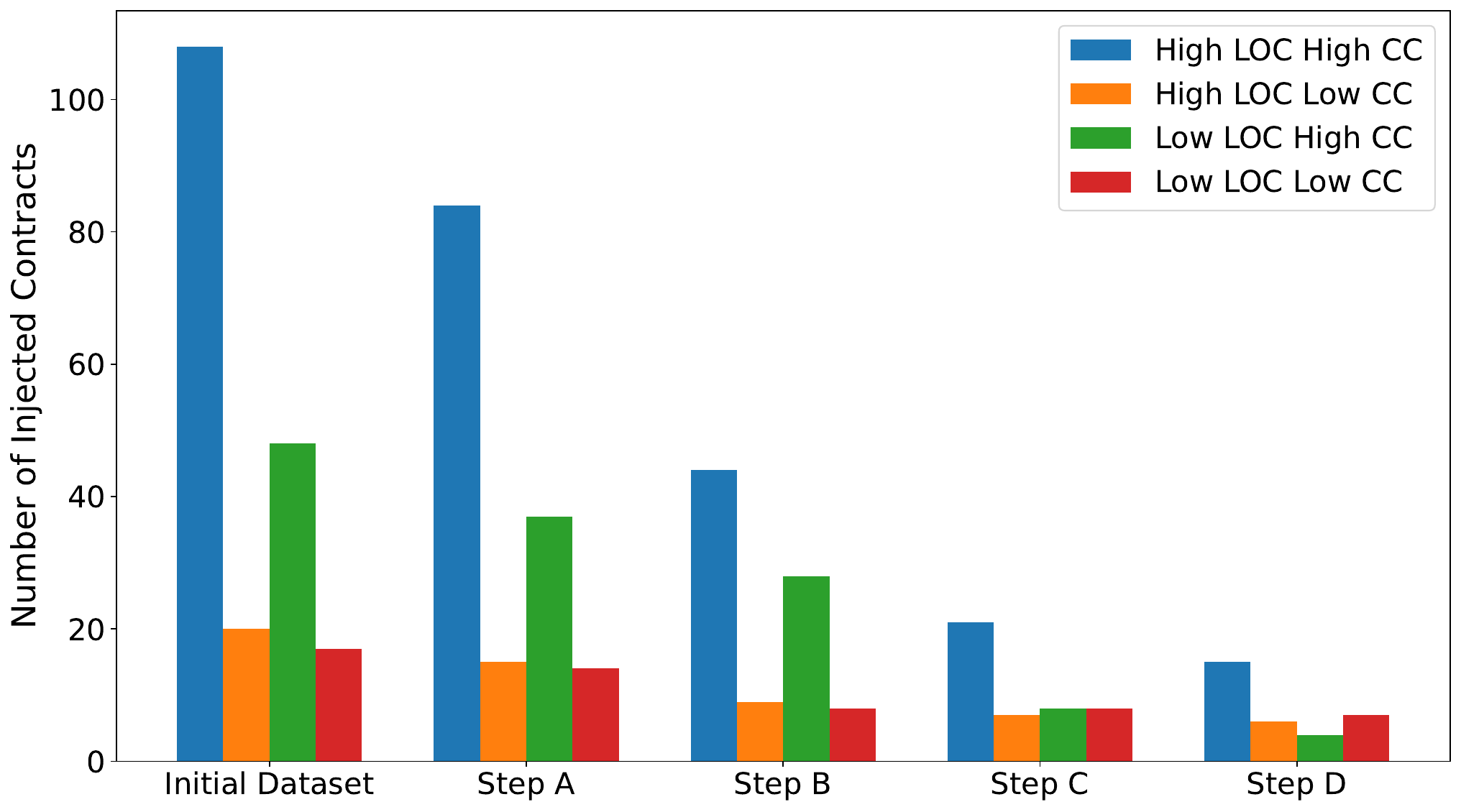}
    \caption{Validation filtering by contract complexity class}    \label{fig:progressive_filtering_contract_complexity}
\end{figure}

To better understand the reduction observed across the validation pipeline, we analyzed whether survival is associated with the structural complexity of the target contracts and the semantic complexity of the injected vulnerabilities. Figure \ref{fig:progressive_filtering_contract_complexity} shows the progression of injected contracts across validation steps for different complexity classes. 
High/Low classes were assigned by median split on LOC and CC.
Simpler contracts tend to survive more frequently, whereas complex contracts undergo a stronger reduction. In particular, contracts in the Low LOC--Low CC class exhibit the highest survival rate, while those in the High LOC--High CC class show the strongest reduction. This suggests that structural complexity, and especially cyclomatic complexity, makes it harder to inject vulnerabilities while preserving syntactic correctness and business logic. This effect is particularly evident in Step~C, which consistently acts as the main bottleneck across complexity classes.

To better understand the role of vulnerability semantics, we grouped the 25 vulnerability types that passed Phase~II according to the ODC defect classification scheme \cite{chillarege1992orthogonal}, according to the mapping in the OpenSCV taxonomy. Table~\ref{tab:defect_type_phase2} shows the resulting distribution. The Checking class dominates, with 11 types, followed by Timing/Serialization (5) and Algorithm/Method (4). The remaining ODC classes cover only a few types each. This distribution reflects that Checking-related vulnerabilities, such as missing input validation or access control weaknesses, are more readily injectable through local edits, while more structural defect types require changes that are harder to reconcile with the injection constraints.

\begin{table}[tbp]
    \centering
    \caption{Vulnerability types passing Phase~II by defect type}
    \label{tab:defect_type_phase2}
    \renewcommand{\arraystretch}{1.1}
    \setlength{\tabcolsep}{5pt}
    \scriptsize
    \resizebox{0.22\textwidth}{!}{
        \begin{tabular}{lc}
            \toprule
            \textbf{Defect Type}      & \textbf{Categories} \\
            \midrule
            Checking                  & 11                  \\
            Timing/Serialization      & 5                   \\
            Algorithm/Method          & 4                   \\
            Function/Class/Object     & 2                   \\
            Assignment/Initialization & 2                   \\
            Interface/O-O Messages    & 1                   \\
            \bottomrule
        \end{tabular}
    }
\end{table}

\begin{figure}[tbp]
    \centering
    \includegraphics[width=\columnwidth]{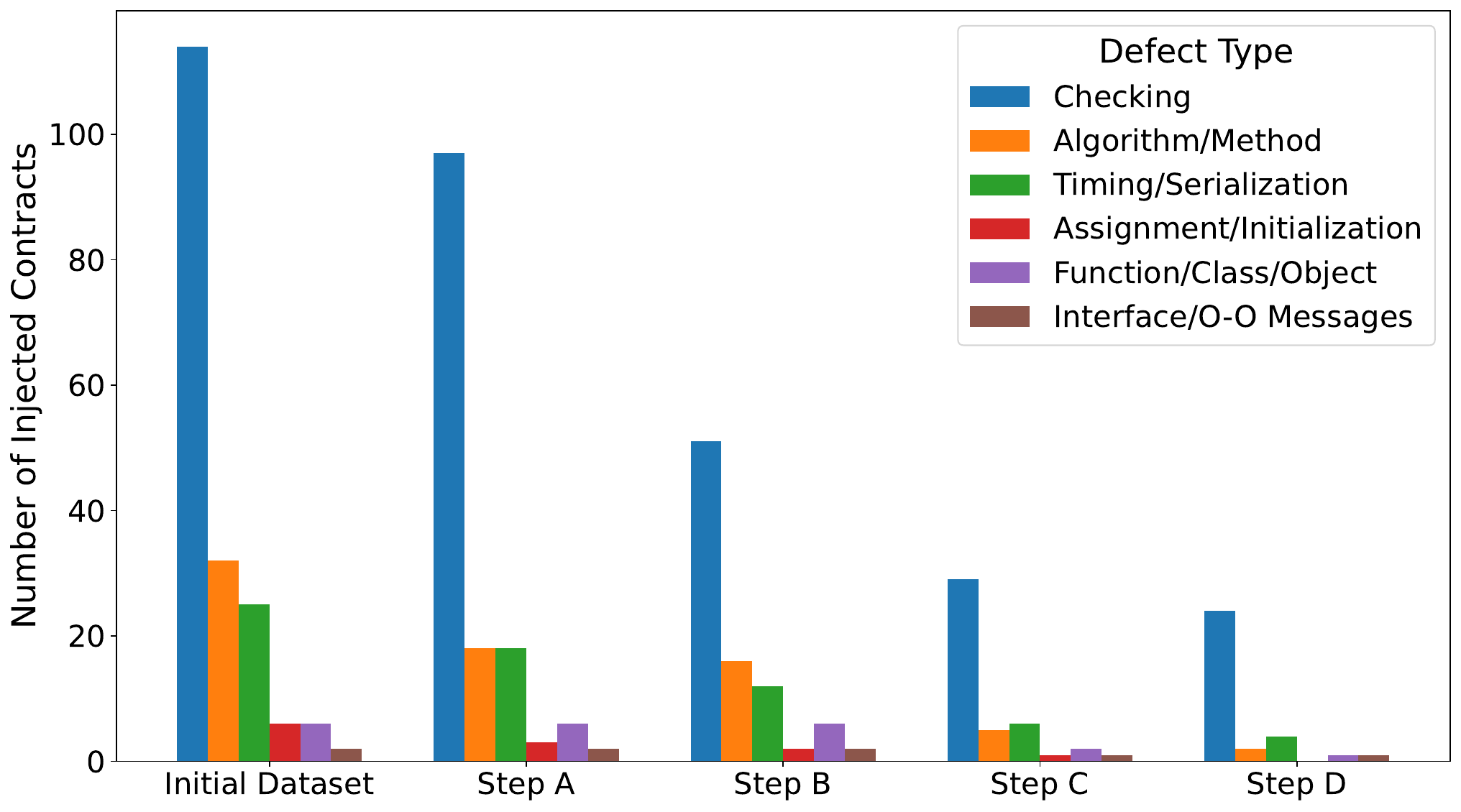}
    \caption{Validation filtering by defect type}
    \label{fig:vuln_defect_filtering}
\end{figure}

A similar pattern emerges when grouping vulnerabilities by defect type (shown in Figure \ref{fig:vuln_defect_filtering}). Vulnerabilities belonging to localized classes, such as Checking, tend to survive the validation pipeline more often, whereas more structural types, such as Algorithm/Method or Assignment/Initialization, show much lower survival. This indicates that injection success depends not only on the contract's complexity but also on the vulnerability pattern's intrinsic complexity.

High injectability predicted in Phase~II does not necessarily imply high survival in Phase~III. Some highly injectable types are later filtered out because they alter business logic or fail to produce semantically meaningful defects. Categories such as Wrong Caller Identification (8.1.1), for which the assessment model achieves near-perfect accuracy \rev{(Fig.~\ref{fig:accuracy_per_vuln}),} also show high survival rates in Phase~III. Conversely, lower-accuracy types tend to show lower survival. This gap indicates that injectability and validation capture different aspects of the process, making a positive assessment necessary but not sufficient for producing a valid vulnerable contract.

\finding{\textbf{Answer to RQ2:} Of 193 non-duplicate injected contracts, 32 (16.58\%) survive the full validation pipeline. Business logic verification (Step~C) is the main bottleneck, reducing survivors from 89 to 44. The defect types most suitable for injection are those in the Checking class (e.g., missing access control, missing input validation), which account for 11 of the 25 types passing Phase~II and achieve the highest survival rates in Phase~III. More structural types, such as Algorithm/Method or Assignment/Initialization, show lower survival because they require modifications harder to reconcile with the injection constraints. This creates a systematic bias: the validated ground truth is skewed toward checking-related vulnerabilities, and taxonomy coverage narrows substantially as contracts progress through the pipeline. Consumers should account for this distribution when drawing conclusions about tool effectiveness.}

\subsection{\revb{Using the Dataset to Compare Detection Tools}}
\label{subsec:demonstration}
\revb{The case study closes with a demonstration of the utility of the validated dataset, which we use to analyze the detection capabilities of existing SATs and thereby answer RQ3. The contracts validated in Step~D are analyzed with the three selected tools, and the tool outputs are compared against the manually established ground truth. Detections are classified as true positives (TP), false positives (FP), false negatives (FN), or static analysis failures (SAF). A \textit{true positive} (TP) corresponds to a correctly detected injected vulnerability that is also manually validated as correct; a \textit{false positive} (FP) indicates a vulnerability reported by the tool but not confirmed during manual analysis; a \textit{false negative} (FN) represents a vulnerability that is present but not detected by the tool; finally, a \textit{static analysis failure} (SAF) denotes cases where the analyzer fails to produce any output.}

\revb{We present this demonstration as a qualitative use of the dataset rather than a full evaluation of the tools.} The metrics below illustrate that the dataset can expose differences between tools; given the small analyzable set (19 contracts), they should be read as indicative observations, not as a tool ranking.
Of 32 validated contracts, 13 could not be analyzed by any of the tools considered and were therefore classified as static analysis failures (SAF). These contracts were retained in the dataset to preserve diversity during manual validation, as previously discussed. The remaining 19 contracts were used for tool-level evaluation.



The comparison reveals that the three tools exhibit partially overlapping detection capabilities. Several verified vulnerabilities are detected by more than one tool, but no tool achieves complete coverage of the validated dataset. Table \ref{tab:static_analysis_results} reports the detection results and aggregate metrics for each tool. Given the evaluation uses 19 contracts, precision and recall should be seen as indicative, not conclusive; small changes (e.g., one TP or FP) can substantially shift the metrics. Slither achieves perfect precision, producing no false positives, but its recall is lower than that of the other tools. Solhint provides the highest recall, while Remix achieves comparable recall at the cost of more false positives. Overall, no single tool provides both perfect precision and high coverage, which suggests that the generated dataset captures vulnerabilities with heterogeneous detection difficulty and that combining multiple analyzers is necessary to improve overall coverage.

\begin{table}[tbp]
    \centering
    \caption{Static analysis results on validated contracts}
    \label{tab:static_analysis_results}
    \renewcommand{\arraystretch}{1.1}
    \setlength{\tabcolsep}{4pt}
    \scriptsize
    \resizebox{0.35\textwidth}{!}{
        \begin{tabular}{lccccccc}
            \toprule
            \textbf{Tool} & \textbf{TP} & \textbf{FP} & \textbf{FN} & \textbf{SAF} & \textbf{Precision} & \textbf{Recall} & \textbf{FDR} \\
            \midrule
            Remix         & 11          & 3           & 5           & 13           & 0.786              & 0.688           & 0.214        \\
            Slither       & 11          & 0           & 8           & 13           & 1.000              & 0.579           & 0.000        \\
            Solhint       & 12          & 2           & 5           & 13           & 0.857              & 0.706           & 0.143        \\
            \bottomrule
        \end{tabular}
    }
\end{table}

Of the 25 types in the dataset, only 9 have at least one confirmed detection. Most true positives concentrate in Wrong Caller Identification (8.1.1), the category with the clearest syntactic signature, with remaining detections spread across types such as Unsafe Credit Transfer (1.1.1) and Missing Verification for Program Termination. Several types surviving Step~D yield only false negatives or static analysis failures. Remix and Solhint show similar category-level coverage, while Slither detects fewer types but produces no false positives. This confirms that static analysis effectiveness is not uniform and that tool combination is necessary to improve recall.

A substantial proportion of failures is due not to the injected vulnerabilities themselves, but to limitations of the tools on legacy Solidity code. The 13 SAF contracts originate from only four target contracts (IDs 2893, 4488, 2854, and 6224), all of which use legacy Solidity versions (0.3.x, 0.4.6, and 0.4.11), while all analyzable contracts use compiler version 0.4.18 or later. The failing contracts also rely on deprecated constructs such as \texttt{sha3}, legacy \texttt{delegatecall}, \texttt{callcode}, and unchecked \texttt{send}. Importantly, running the same tools on the original, unmodified versions of these contracts produces the same errors, confirming that the SAF phenomenon reflects a structural limitation of current tools on legacy code rather than a weakness of the injection \revb{approach}.

Phase~II demonstrates that LLMs can identify plausible contract--vulnerability pairs and generate candidate injections, while Phase~III reveals that only a limited subset satisfies all validation criteria. The progressive reduction across the pipeline shows that syntactic correctness alone is insufficient, and that preserving business logic and semantic validity are the main challenges. The comparison with SATs further shows that current analyzers provide incomplete and heterogeneous coverage, and may fail entirely on legacy code.

\finding{\textbf{Answer to RQ3:} The generated dataset reveals clear differences in tool behavior. The three tools show distinct detection profiles: Slither achieves perfect precision but lower recall (57.9\%), Solhint achieves the highest recall (70.6\%) with some false positives, and Remix falls between the two. Their detection coverage is partially complementary, with each tool catching vulnerabilities that the others miss. Detection success also varies by vulnerability type: only 9 of the 25 injected types are detected by at least one tool, and most true positives concentrate in a few types with clear syntactic signatures. These patterns would not be visible without a dataset containing diverse, verified vulnerable contracts.}

We examine the overall injection process further in two ways. First, to check whether model size or context length changes the picture, we used a long-context proprietary model (Claude Sonnet 5, 1M-token context window) as the injector on three contracts (330, 1472, and 5330), under the same constraints and with the candidates validated by the human evaluator. Of 25 candidates, 9 (36\%) passed all steps, concentrating in the same localized types the smaller model favored (Wrong Caller Identification, Typographical Error, unchecked low-level calls); value-transfer, transaction-order, reentrancy, and arithmetic variants were again rejected. Producing a correctly typed, exploitable defect while preserving business logic is thus intrinsic to the task, not a matter of model size or context. We treat this as a complementary proprietary-model probe; the main pipeline stays open-source, and rates are not directly comparable given three simple contracts and no deduplication.

Second, we compare our approach against SolidiFI~\cite{ghaleb2020issta}, a pattern-based injector. It applied to only 3 of our 14 contracts (330, 1472, and 5330); on the other 11 it failed at compilation, its toolchain pinned to a single compiler (Solidity 0.5.12). This is the same unsupported-version limitation we report for the analysis tools (Section~\ref{sec:lessons}). SolidiFI is also restricted to seven fixed bug types, and supporting new types or compiler versions requires porting effort; our LLM-based approach instead targets the 49 OpenSCV types and is not tied to a specific compiler version. On the three where it ran, SolidiFI injected 110 candidates, dozens per contract, all by code-snippet insertion, appending self-contained functions and state variables and leaving the original code paths unchanged, unlike the single in-place edit of our pipeline. On \texttt{tx.origin}, overlapping our Wrong Caller Identification, it adds standalone functions checking \texttt{require(tx.origin == owner)} on a parameter while leaving the contract's own access-control modifier untouched; our injection instead rewrites that modifier in place. The injected snippets thus sit alongside the contract instead of within its logic, and tend to not form realistic in-place vulnerable variants. The two methods pursue different goals. SolidiFI seeds detectable patterns for tool evaluation, while ours targets realistic in-place vulnerabilities, so a direct survival-rate comparison is not meaningful.
\section{Lessons Learned and Practical Implications}
\label{sec:lessons}

\textbf{Assessment accuracy does not predict injection success.} The assessment phase identifies contract--vulnerability pairs that appear plausible, yet many fail during validation. The gap arises because the assessment model reasons about local syntactic feasibility, while successful injection also requires preserving business logic and satisfying structural constraints. As shown in Table~\ref{tab:pipeline_summary}, only 32 of 193 non-duplicate contracts survive all steps. \revb{Future injection approaches} should treat assessment as a necessary filter rather than a reliability indicator, and invest in intermediate validation checkpoints.

\textbf{Business logic verification is the main bottleneck.} Step~C causes the largest \revb{relative} drop in the pipeline, reducing survivors from 89 to 44. Many injections that compile and execute correctly still alter the intended contract behavior in ways that are only detectable through semantic analysis. This suggests that improving the injection prompt alone is insufficient; the generation process would benefit from an explicit semantic preservation objective, such as differential testing against the original contract or formal equivalence checks on non-vulnerable execution paths.

\textbf{LLM-based injection converges toward a narrow set of patterns.} The 80\% reduction from 997 raw generations to 193 distinct contracts shows that the model, under fixed constraints, repeatedly produces similar outputs. This limits the diversity of the resulting dataset and means that increasing the number of generation runs yields diminishing returns. We did not explore mitigation strategies such as temperature variation, prompt perturbation, or multi-model ensembles; investigating their effect on injection diversity is a concrete direction for future work. Additionally, the textual deduplication method used introduces a dual-direction approximation error: formatting-only variants are counted as distinct, while textually identical outputs on different execution paths are merged. Future work should consider semantic equivalence checks, e.g., through symbolic execution, as a more robust deduplication criterion.

\textbf{Contract complexity amplifies injection difficulty.} Contracts with higher CC and more external interactions exhibit lower survival rates across all validation steps (shown in \rev{Fig.~\ref{fig:progressive_filtering_contract_complexity}).} This effect is visible in Step~C, where complex control flow makes introducing vulnerabilities without side effects harder. Practitioners should expect lower yields on structurally complex contracts. Possible mitigations include relaxing injection constraints for complex targets (e.g., allowing modifications to more than one function) or breaking the injection into smaller steps.

\textbf{Defect type determines both injectability and survival.} The Checking class accounts for 11 of the 25 types passing Phase~II and achieves the highest survival rate in Phase~III, because checking-related vulnerabilities (e.g., missing access control) can be introduced through minimal local edits. More structural types, such as Algorithm/Method, require more extensive modifications harder to reconcile with the injection constraints. Dataset builders should consider complementary injection strategies for underrepresented defect classes. The ground truth is therefore not uniformly representative of the OpenSCV taxonomy: types harder to inject via local edits are systematically underrepresented. Tool evaluations on this dataset measure coverage primarily over the Checking class and should not be generalized to other defect types.

\textbf{Manual validation is a scalability bottleneck.} Steps~B, C, and~D require manual inspection of each injected contract, which limits the number of contracts that can be processed in practice. In \revb{the case study}, a single evaluator inspected all candidates, and the effort grew linearly with the number of surviving contracts at each step. For the \revb{approach} to scale to larger contract sets or to be applied repeatedly with different LLMs, parts of the manual validation would need to be automated. Step~C (business logic verification) is the target \rev{that lends itself best to} automation, for instance, through differential testing that compares the behavior of the original and injected contracts under a set of representative inputs. Steps~B and~D, which involve semantic judgment about constraint compliance and exploitability, are harder to automate but could be supported by LLM-assisted review.

\textbf{SATs fail on legacy code.} Of the 32 validated contracts, 13 (40.6\%) could not be analyzed by any of the three tools due to static analysis failures. The root cause is not the injected vulnerability itself but the use of legacy Solidity versions (0.3.x--0.4.11) and deprecated constructs. Running the same tools on the original, unmodified contracts produces the same errors. Additionally, any evaluation that relies on static analysis over real-world Ethereum contracts must account for deployed code falling outside the supported range of current analyzers. Legacy versions also account for a large share of the failures during injection itself: 26 of the 43 Step~A compilation failures arise because the injector introduces constructs unsupported by the contract's compiler (Section~\ref{sec:results}).

These lessons translate into a few \textbf{practical takeaways} for practitioners building vulnerable-contract datasets, as the same difficulties recur across models. Injectability assessment should be treated as a filter rather than a guarantee, with intermediate validation checkpoints, \rev{because} business-logic preservation, not compilation, is where most candidates fail. Manual validation dominates cost and is worth planning for, with the business-logic check the first candidate for automation through differential testing. Lower yields should be expected on structurally complex contracts and on defect types outside the Checking class, so the resulting datasets are skewed and should not be used to draw tool-coverage conclusions beyond that class. Adding generation runs does not improve diversity, as outputs converge; varying temperature, perturbing prompts, or combining models may be more effective. Finally, legacy Solidity contracts are a frequent source of both injection and static-analysis failures. These takeaways assume a single trained evaluator applying predefined criteria and open-source quantized models on constrained hardware, and are bounded by a small validated set and a defect-type distribution skewed toward Checking-class vulnerabilities.
\section{Threats to Validity}
\label{sec:threats}

\textbf{Internal validity.} Steps~B, C, and~D \rev{of Phase~III} rely on manual inspection by a single evaluator, which may introduce \rev{subjectivity, and inter-rater agreement was not measured. We mitigated this by applying predefined acceptance and rejection criteria consistently at each step (Section~\ref{subsec:Phase2}), and by re-examining borderline cases deliberately, although full double-annotation was not feasible within the scope of this study.} The 50\% injectability threshold and the 10 repeated runs per pair are pragmatic \rev{decisions that influence which types are retained and how stable the injectability scores are.}

\textbf{Construct validity.} The injectability score is a proxy \rev{for true injection feasibility,} based on repeated LLM predictions rather than formal \rev{structural analysis. As Fig.~\ref{fig:confusion_matrix_manual_validation} shows, this proxy produces both false positives and false} negatives, so the 25 retained types may underestimate the \rev{space of injectable vulnerabilities. Duplicate removal works by textual comparison after whitespace normalization, which means that semantically equivalent injections differing only in non-whitespace formatting count as distinct, while textually identical injections affecting different execution paths are merged.

\textbf{External validity.} \revb{The case study} rests on 14 contracts from the SmartBugs dataset, so the results may not generalize to larger or more diverse contract sets. \revb{It} also relies on two specific open-source models run with quantization} on constrained hardware, and different models or prompting strategies could \rev{yield different outcomes. Closed-source models (e.g., GPT-4, Gemini)} were excluded from the main pipeline for reproducibility and \rev{because per-query costs would limit large-scale experimentation. The model selection for the injection phase was validated on a single vulnerability type (Integer Underflow/Overflow), and the relative performance of models may differ for other defect classes.}

\textbf{Conclusion validity.} The reliability check covers 108 of the 6,860 assessment \rev{decisions (14 contracts $\times$ 49 types $\times$ 10 runs), so the observed accuracy may not hold uniformly across types. The} 95\% \rev{Wilson confidence interval for the observed} 75.9\% \rev{accuracy is approximately} [67.3\%, 83.2\%], \rev{which leaves non-negligible uncertainty. The tool analysis is limited by the} 13 static analysis failures on legacy Solidity code, which reduce the analyzable set from 32 to 19 \rev{contracts. At this scale, single-contract differences in ground truth classification would materially shift precision and recall, so the reported tool metrics should be read as exploratory rather than definitive.}

\section{Conclusion}

LLM-based vulnerability injection into Solidity smart contracts is feasible but \rev{far from straightforward. Of the} 193 non-duplicate injected contracts, only 32 (16.58\%) passed \rev{every validation step, and preserving the original business logic was the main bottleneck. Injection success depends on both contract complexity and defect type, with structurally simpler contracts and vulnerability types that map to localized syntactic patterns reaching the highest survival rates. We obtained these results with \revb{the proposed approach, which combines injectability assessment, LLM-based injection, and a multi-step validation pipeline}. Applying the resulting contracts to three static analysis tools showed that current analyzers cover the injected defects incompletely and unevenly, and that they fail outright on legacy Solidity} code.

Future work will extend the evaluation to \rev{a larger and more diverse set of contracts, and will explore alternative models and prompting strategies, including closed-source ones, to widen injection diversity and reduce semantic errors. The diversity ceiling that follows from model convergence deserves particular attention, because additional generation runs do little to raise it. Automating parts of the manual validation, for instance through differential testing or semantic deduplication, would let the pipeline scale further.}

\revb{\section*{Acknowledgments}
This work is funded by national funds through FCT -- Foundation for Science and Technology, I.P., within the scope of the research unit UID/00326 -- Centre for Informatics and Systems of the University of Coimbra, \url{https://doi.org/10.54499/UID/00326/2025}.}

\bibliographystyle{IEEEtran}
\bibliography{bibliography}

\end{document}